\documentclass[preprint,amsmath,amssymb,aps,superscriptaddress,nofootinbib,prd]{revtex4-1}
\usepackage{graphicx}
\usepackage{xcolor}

\begin{document}

\title{Ultra-relativistic freeze-out of fermion dark matter through a heavy scalar mediator}

\author{Jin-Lei Yang}\email{jlyang@hbu.edu.cn}
\author{Ke-Shuo Cai}
\author{Yin-Jie Zhang}
\author{Hai-Bin Zhang}
\author{Tai-Fu Feng}
\affiliation{Department of Physics, Hebei University, Baoding, 071002, China}
\affiliation{Key Laboratory of High-precision Computation and Application of Quantum Field Theory of Hebei Province, Baoding, 071002, China}
\affiliation{Research Center for Computational Physics of Hebei Province, Baoding, 071002, China}

\date{\today}

\begin{abstract}
We investigate the production of Majorana fermion dark matter (DM) via a heavy scalar mediator during reheating, with particular emphasis on ultra-relativistic freeze-out (UFO). Depending on the mediator mass, interaction strengths, and reheating temperature, we identify three distinct production regimes: freeze-in, WIMP-like freeze-out, and UFO, all of which can account for the observed DM relic abundance. In the UFO regime, significant DM production can persist after chemical decoupling, while entropy injection during reheating simultaneously dilutes the DM yield. An accurate determination of the final relic abundance therefore requires a consistent treatment of both continued production and entropy dilution. We further examine the impact of scalar cubic interactions and find that they modify the boundaries between the different production regimes and reshape the relic-density contours. In particular, the results are more sensitive to the cubic coupling involving the mediator and the light scalar than to the heavy-scalar self-coupling.
\end{abstract}

\maketitle


\section{Introduction}
\label{sec:intro}

The Standard Model (SM) provides a remarkably successful description of elementary particles and their interactions. However, it contains no viable particle candidate for the dark matter (DM) inferred from astrophysical and 
cosmological observations. Evidence for DM comes from, among other observations, galaxy rotation curves~\cite{Sofue:2000jx} and gravitational lensing~\cite{Clowe:2006eq,Metcalf:2003sz,Bartelmann:2010fz}. Cosmological observations indicate that DM contributes approximately one quarter of the present-day cosmic energy density~\cite{WMAP:2012nax,Planck:2018vyg}. In a particle interpretation, DM is generally expected to be electrically neutral and colorless, but its microscopic nature and production mechanism remain unknown. Measurements of the cosmic microwave background (CMB) constrain the DM abundance with high precision~\cite{Planck:2018vyg}:
\begin{equation}
    \Omega_{\mathrm{DM}}h^2 = 0.120 \pm 0.001.
    \label{eqomega}
\end{equation}
The relation between the present-day DM relic abundance and its microscopic interactions depends on the thermal history of the early Universe, particularly during the period preceding radiation domination. DM relic-abundance calculations can thus constrain particle interactions and the thermal history of the early Universe.

Two widely studied production mechanisms are thermal freeze-out (FO)~\cite{Arcadi:2017kky,Roszkowski:2017nbc,Steigman:2012nb} and freeze-in~\cite{McDonald:2001vt,Hall:2009bx,Elahi:2014fsa,Bernal:2017kxu}. In the conventional weakly interacting massive particle (WIMP) scenario, DM first attains chemical equilibrium with the thermal bath. As the Universe cools, number-changing reactions eventually fail to maintain equilibrium, leading to nonrelativistic freeze-out. In freeze-in, feeble interactions populate DM without establishing chemical equilibrium with the bath. Such candidates are commonly referred to as feebly interacting massive particles (FIMPs). Both mechanisms have been studied in many extensions of the SM. Representative candidates include the lightest neutralinos in supersymmetric (SUSY) models~\cite{Jungman:1995df,Edsjo:1997bg,Ellis:1998kh,Feng:2000gh,Ellis:2003cw,Arkani-Hamed:2006wnf,Cheung:2012qy,Cohen:2013ama,Huang:2014xua,Han:2014nba,Badziak:2015exr,Beneke:2016ync,Olive:2016efh,GAMBIT:2017zdo,Drees:2018dsj,Han:2019vxi,Chakraborti:2021kkr,Barman:2022jdg,Yang:2023krd,Bisal:2023iip,Martin:2024ytt,Antusch:2025rbp,Yue:2025wnm,Bhattiprolu:2025zwt,Khan:2025ibo,Hamaguchi:2026fou}, as well as scalar singlet
DM~\cite{McDonald:1993ex,Burgess:2000yq,He:2008qm,He:2009yd,Gonderinger:2009jp,Profumo:2010kp,Guo:2010hq,Bandyopadhyay:2010cc,Yaguna:2011qn,Cline:2013gha,Feng:2014vea,GAMBIT:2017gge,Athron:2018ipf,Mandal:2019oth,Beck:2021xsv,Basak:2021tnj,Qi:2021rhh,Coleppa:2023vfh,Avila:2025qsc} and gauge-singlet fermionic
DM~\cite{Kim:2006af,Kim:2008pp,Baek:2011aa,Lopez-Honorez:2012tov,Baek:2012uj,Klasen:2013ypa,Esch:2013rta,Matsumoto:2014rxa,Fedderke:2014wda,Molinaro:2014lfa,GAMBIT:2018eea,Arias:2020qty,Yaguna:2023kyu,Das:2026zuo} in non-SUSY extensions of the SM. 

WIMP-like DM candidate is well-motivated and can naturally account for the observed value of $\Omega_{\rm DM}h^2$, but the  corresponding parameter space is pushed into a tight corner experimentally~\cite{PandaX:2024qfu,LZ:2024zvo,XENON:2025vwd,LZ:2025igz,XENON:2026gxb} due to its relatively large DM-SM interactions. Although FIMPs can evade these constraints, their feeble couplings to the SM make them challenging to probe through direct or indirect detection~\cite{Bernal:2017kxu}. An interesting alternative is ultra-relativistic freeze-out (UFO)~\cite{Henrich:2025gsd,Henrich:2025sli}, in which DM first reaches chemical equilibrium with the thermal bath and subsequently decouples at a temperature well above its mass. For suitable couplings and reheating histories, UFO may occur between the conventional freeze-in and WIMP-like freeze-out regimes. This connection has been explored in a model with a heavy vector mediator~\cite{Henrich:2025pca}. The direct detection prospects of UFO have been studied in Ref.~\cite{Henrich:2026tox}, while the effects of gravitational particle production on its parameter space have been investigated in Ref.~\cite{Gross:2026bhe}. In our previous work~\cite{Yang:2026syo}, we proposed a CP-odd scalar DM candidate and found that UFO can account for its relic abundance over an extended viable parameter region.

Since additional scalar fields arise in many extensions of the SM, the realization of UFO in a heavy-vector portal~\cite{Henrich:2025pca} raises the question of whether the mechanism persists for scalar mediated interactions. Heavy scalar mediators provide a natural setting for this investigation. The realization of UFO depends on the temperature dependence of number-changing interaction rates relative to the cosmic expansion rate. Scalar and vector mediators can produce different temperature dependence of the number-changing rates, and hence different decoupling histories. We therefore investigate the production of Majorana fermion DM through a heavy scalar mediator.\footnote{For the case of Dirac DM, the absence of identical-particle factor doubles the production rates and the relic abundance, while the normalized abundance evolution approximately unchanged. The qualitative conclusions obtained in this work are therefore expected to remain applicable.} By identifying the conditions for chemical equilibration and subsequent decoupling, we assess whether scalar interactions can support UFO and clarify how the underlying particle interactions govern the connection between WIMP-like FO, FI, and UFO.
    
The paper is organized as follows. In Sec.~\ref{sec:model}, we introduce the model and present its scalar potential, mass spectrum, and relevant interactions. We derive the DM production rates in Sec.~\ref{sec:DMrate}. The cosmological evolution and the calculation of the DM relic abundance from the Boltzmann equation are described in Sec.~\ref{sec:boltzmann}. We discuss the numerical results in Sec.~\ref{sec:numerical} and present our conclusions in Sec.~\ref{sec:conclusions}. The tedious formula are collected in the appendix.


\section{The Model}
\label{sec:model}

We extend the SM by a real gauge-singlet scalar $\phi$ and a Majorana fermion $\chi$. In the absence of a symmetry that fixes the origin of $\phi$, we shift the singlet field so that its vacuum expectation value (VEV) vanishes. This redefinition changes the coefficients in the scalar potential but not its physical content. The scalar potential of the model is
\begin{align}
    V=&-\frac{1}{2}\mu^2 H^\dagger H+\lambda(H^\dagger H)^2+\mu_\phi^3\phi+\frac{1}{2}m_\phi^2 \phi^2+\frac{1}{6}\kappa \phi^3+\kappa_{H\phi}\phi H^\dagger H+\frac{1}{4}\lambda_\phi\phi^4\nonumber\\
    &+\frac{1}{2}\lambda_{H\phi}\phi^2H^\dagger H,
\label{potential}
\end{align}
where $H$ is the SM Higgs doublet, and $H=(0,(v+h)/\sqrt2)^T$ with $v\simeq246~\mathrm{GeV}$. The tadpole conditions at $\langle h\rangle=\langle\phi\rangle=0$ give
\begin{align}
    \mu^2=2\lambda v^2,\quad \mu_\phi^3=-\frac{1}{2}\kappa_{H\phi}v^2.
    \label{potentialbound}
\end{align}
Using these conditions, the mass-squared matrix of the CP-even scalars in the $(h,\phi)$ basis is
\begin{align}
    m_h^2=
    \begin{pmatrix}
        2\lambda v^2
        & \kappa_{H\phi}v \\[4pt]
        \kappa_{H\phi}v
        & m_\phi^2+\frac{1}{2}\lambda_{H\phi}v^2
    \end{pmatrix}.
    \label{Higgsmassmatrix}
\end{align}
For the electroweak stationary point to be a local minimum, the scalar mass squared matrix must be positive definite, as required by
\begin{align}
    \lambda>0,\quad m_\phi^2+\frac{1}{2}\lambda_{H\phi}v^2>\frac{\kappa_{H\phi}^2}{2\lambda}.
\end{align}
In the heavy-mediator limit, with $m_\phi^2\gg v^2$ and $|\kappa_{H\phi}|v\ll m_\phi^2$, the masses of the physical scalar states $h_1$ and $h_2$ are approximately
\begin{align}
    &M_{h_1}^2\approx 2\lambda v^2-\frac{\kappa_{H\phi}^2}{m_\phi^2}v^2,\nonumber\\
    &M_{h_2}^2\approx m_\phi^2+\left(\frac{\kappa_{H\phi}^2}{m_\phi^2}+\frac{1}{2}\lambda_{H\phi}\right)v^2.
    \label{physicalHiggsmass}
\end{align}

The mixing between the doublet and singlet scalar fields is characterized by the angle $\theta_{\rm mix}$. In the heavy-mediator limit with small mixing,
\begin{align}
    \sin^2\theta_{\rm mix}\approx \frac{\kappa_{H\phi}^2 v^2}{m_\phi^4}.
    \label{mixingangle}
\end{align}
For the representative benchmark $\kappa_{H\phi}=100~\mathrm{GeV}$ and $m_\phi=10^4~\mathrm{GeV}$, this gives $\sin^2\theta_{\rm mix}\simeq 10^{-8}$. The corresponding shifts in the tree-level couplings of the SM-like Higgs boson to SM particles are therefore small. Scalar mixing also induces interactions between DM and SM fermions through scalar exchange. The suppression of these interactions by the small mixing angle can alleviate constraints from direct detection~\cite{PandaX-II:2016vec,LUX:2016ggv,XENON:2017vdw,PICO:2017tgi}, indirect detection~\cite{Giesen:2015ufa,MAGIC:2016xys,HESS:2016mib,DelaTorreLuque:2024ozf,Fermi-LAT:2025gei}, and collider searches~\cite{ATLAS:2015wrn,CMS:2016lcl,ATLAS:2023tkt,CMS:2023sdw}. The Majorana mass and singlet Yukawa interaction are given as
\begin{align}
    \mathcal{L}_\chi=-\frac{1}{2} M_\chi \bar\chi \chi-\frac{1}{2} y_\chi\phi\bar\chi \chi,
    \label{DMcoupling}
\end{align}
with $\chi$ a four-component Majorana field and $y_\chi$ 
the dimensionless Yukawa coupling.


\section{The DM production rate}
\label{sec:DMrate}

\begin{figure*}[t!]
\centering
    \includegraphics[width=1.\textwidth]{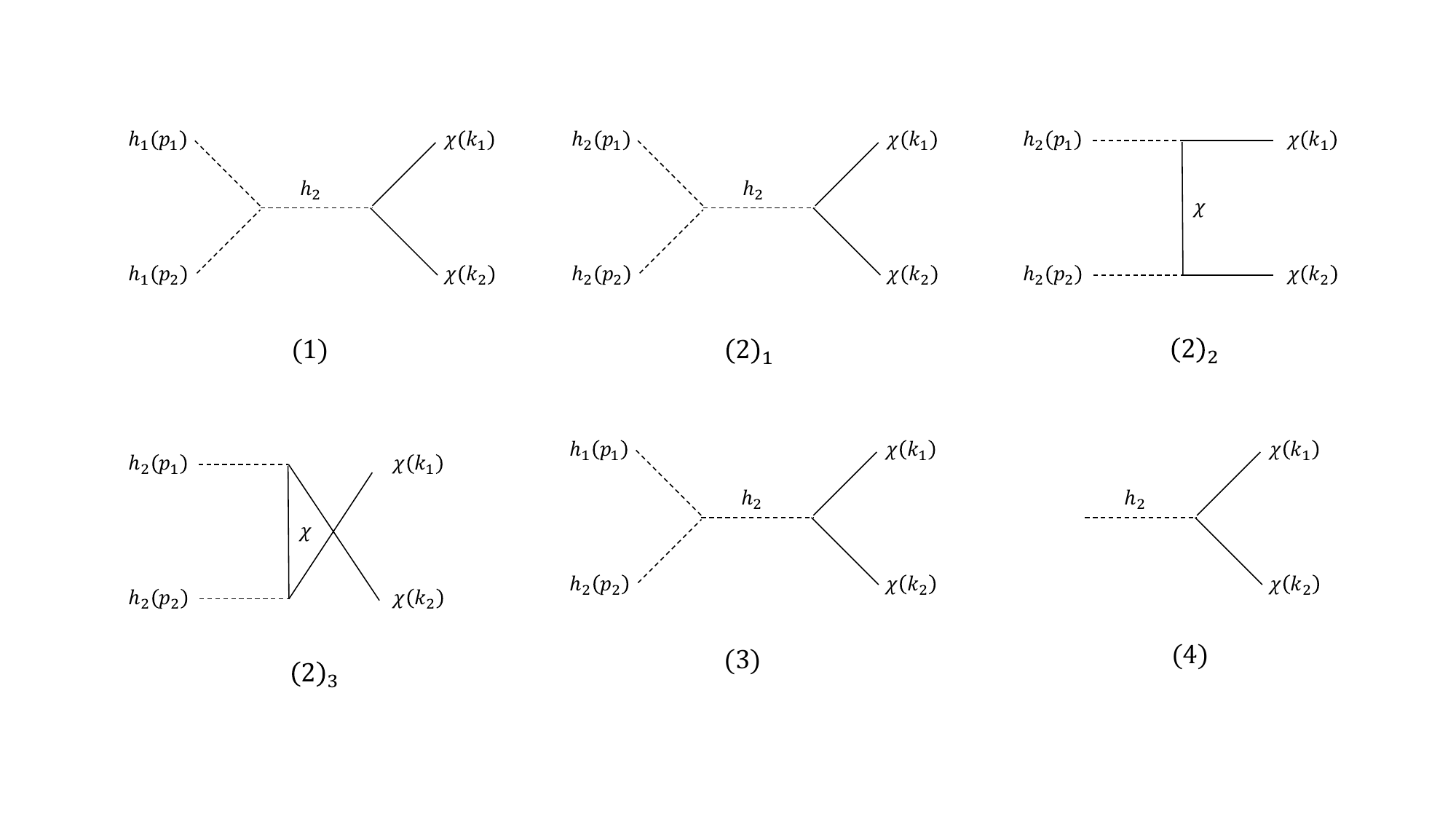}
    \caption{The dominant Feynman diagrams contributing to the DM production rate.}
\label{figrate}
\end{figure*}

In this section, we neglect scalar-mixing corrections while retaining the direct scalar interaction vertices as a simplified example to illustrate how to calculate the DM production rate. The relevant diagrams are shown in Fig.~\ref{figrate}, which are used to derive the amplitudes, cross sections, decay widths and thermal reaction densities in this section. The complete calculation, including scalar mixing, is given in the appendix. The full production rate used in our analysis additionally includes scalar-mixing effects and the $W^+W^-$, $ZZ$, and $t\bar t$ initial states, as specified below.

Figs.~\ref{figrate} $(1)$, $(2)_{1,2,3}$, $(3)$, $(4)$ correspond to the processes $h_1h_1\to\chi\chi$, $h_2h_2\to\chi\chi$,
$h_1h_2\to\chi\chi$ and $h_2\to\chi\chi$, respectively. The decay channel is kinematically allowed when $M_{h_2}>2M_\chi$.
For the processes $h_i(p_1)h_j(p_2)\rightarrow\chi(k_1)\chi(k_2)$,
the Mandelstam variables are defined as
\begin{align}
s=(p_1+p_2)^2,\;
t=(p_1-k_1)^2,\;
u=(p_1-k_2)^2,\;
s+t+u=M_{h_i}^2+M_{h_j}^2+2M_\chi^2.
\label{eq:production_mandelstam}
\end{align}
where the last equation follows from the on-shell conditions for the external particles. Up to a common overall phase, the $s$-channel amplitude is
\begin{align}
\mathcal{M}_{h_ih_j}^{(s)}=\frac{y_\chi C_{h_ih_jh_2}}{s-M_{h_2}^2+iM_{h_2}\Gamma_{h_2}}
\overline{u}(k_2)v(k_1).
\label{s_channel_amplitude}
\end{align}
Here $C_{h_ih_jh_2}$  is the trilinear scalar vertex coefficient and $\Gamma_{h_2}$ is the total width of $h_2$. Neglecting scalar mixing, the potential in Eq.~\eqref{potential} gives 
\begin{align}
C_{h_1h_1h_2}=\kappa_{H\phi},\;C_{h_2h_2h_2}=\kappa,\;C_{h_1h_2h_2}=\lambda_{H\phi}v
\label{coupling_constants}
\end{align}
for the $s-$channel contributions to $h_1h_1\rightarrow\chi\chi$, $h_2h_2\rightarrow\chi\chi$, and $h_1h_2\rightarrow\chi\chi$, respectively.  

For the process $h_2h_2\to\chi\chi$, fermion exchange also contributes through the $t$ and $u$ channels. Expressing both contributions with the same ordering of the external spinors and using their on-shell equations gives
\begin{align}
\mathcal{M}_{h_2h_2}^{(t)}&=\frac{y_\chi^2}{t-m_\chi^2}\overline{u}(k_2)
\left(\not p_1+2M_\chi\right)v(k_1),
\\
\mathcal{M}_{h_2h_2}^{(u)}&=\frac{y_\chi^2}{u-m_\chi^2}\overline{u}(k_2)
\left(-\not p_1+2M_\chi\right)v(k_1).
\label{eq:scalar_tu_amplitudes}
\end{align}
Consequently,
\begin{align}
\mathcal{M}_{h_1h_1}=\mathcal{M}_{h_1h_1}^{(s)},\;
\mathcal{M}_{h_1h_2}=\mathcal{M}_{h_1h_2}^{(s)},\;
\mathcal{M}_{h_2h_2}
=\mathcal{M}_{h_2h_2}^{(s)}
+\mathcal{M}_{h_2h_2}^{(t)}
+\mathcal{M}_{h_2h_2}^{(u)}.
\label{eq:complete_amplitudes}
\end{align}

Summing over the final-state spins, the pure $s$-channel
contribution is
\begin{equation}
\sum_{\mathrm{spins}}
\left|\mathcal{M}_{h_ih_j}^{(s)}\right|^2
=
\frac{2y_\chi^2 C_{h_ih_jh_2}^2(s-4M_\chi^2)}
     {(s-M_{h_2}^2)^2+M_{h_2}^2\Gamma_{h_2}^2}.
\label{eq:scalar_s_squared}
\end{equation}
The complete squared amplitude for the $h_2h_2$ channel can be expressed by defining
\begin{align}
A(s,t,u)&=\frac{y_\chi C_{h_2h_2h_2}}{s-M_{h_2}^2+i M_{h_2}\Gamma_{h_2}}+2M_\chi y_\chi^2\left(\frac{1}{t-M_\chi^2}+\frac{1}{u-M_\chi^2}\right),
\\
B(s,t,u)&=y_\chi^2\left(\frac{1}{t-M_\chi^2}-\frac{1}{u-M_\chi^2}\right).
\label{eq:scalar_AB}
\end{align}
With $\mathcal{M}_{h_2h_2}=\overline{u}(k_2)(A+B\not p_1)v(k_1)$, one obtains
\begin{align}
\sum_{\mathrm{spins}}|\mathcal{M}_{h_2h_2}|^2={}&2(s-4M_\chi^2)|A|^2+4M_\chi(u-t)\operatorname{Re}(AB^*)\nonumber\\
&+2\left[(M_{h_2}^2+M_\chi^2-t)(M_{h_2}^2+M_\chi^2-u)-sM_{h_2}^2\right]|B|^2.
\label{eq:scalar_22_squared}
\end{align}

The total scattering cross section is
\begin{align}
\sigma_{ij}(s)=\frac{1}{64\pi s}\frac{\sqrt{\lambda(s,M_\chi^2,M_\chi^2)}}{\sqrt{\lambda(s,M_{h_i}^2,M_{h_j}^2)}}\int_{-1}^{1}d\cos\theta\,\sum_{\mathrm{spins}}|\mathcal{M}_{h_ih_j}|^2.
\label{eq:production_cross_section}
\end{align}
The factor $1/2!$ for the identical Majorana fermions in the final state is included in the above equation, and
\begin{align}
\lambda(x,y,z)=x^2+y^2+z^2-2xy-2xz-2yz.
\end{align}
$\theta$ is the angle between $\boldsymbol{p}_1$ and $\boldsymbol{k}_1$ in the center-of-mass frame. For a pure $s$-channel contribution, the angular integral can be performed directly
\begin{align}
\sigma_{ij}^{(s)}(s)=\frac{y_\chi^2 C_{h_ih_jh_2}^2}
     {16\pi [(s-M_{h_2}^2)^2+M_{h_2}^2\Gamma_{h_2}^2]}
\frac{s}{\sqrt{\lambda(s,M_{h_i}^2,M_{h_j}^2)}}
\left(1-\frac{4M_\chi^2}{s}\right)^{3/2}.
\label{eq:scalar_s_cross_section}
\end{align}
In the negligible-mixing approximation, Eq.~\eqref{eq:scalar_s_cross_section} gives the $h_1h_1$ and $h_1h_2$ cross sections, the $h_2h_2$ cross section also includes fermion exchange and interference as given in Eq.~\eqref{eq:scalar_22_squared}.

We evaluate the reaction densities with equilibrium Maxwell–Boltzmann distributions for the bath particles, neglecting the thermal corrections to masses and couplings. Each real scalar has one internal degree of freedom. The scattering reaction density, defined as the number of reactions per unit volume per unit time, is~\cite{Giudice:2003jh}
\begin{align}
\gamma_{ij}(T)=\frac{T}{32\pi^4(1+\delta_{ij})}\int_{s_{ij}^{\min}}^\infty ds\,\frac{\lambda(s,M_{h_i}^2,M_{h_j}^2)}{\sqrt{s}}\,\sigma_{ij}(s)K_1\!\left(\frac{\sqrt{s}}{T}\right),
\label{eq:scattering_reaction_density}
\end{align}
with
\begin{align}
s_{ij}^{\min}=\max\left\{(M_{h_i}+M_{h_j})^2,\,4M_\chi^2\right\}.
\end{align}
Here $K_n$ denotes the modified Bessel function of the second kind. The factor $1/(1+\delta_{ij})$ avoids double counting identical incoming particles and is separate from the final-state symmetry factor already included in $\sigma_{ij}$. Defining the dimensionless reduced cross section as~\cite{Giudice:2003jh}
\begin{align}
\widehat{\sigma}_{ij}(s)=\frac{2\lambda(s,M_{h_i}^2,M_{h_j}^2)}{s}\,\sigma_{h_ih_j}(s),
\label{eq:reduced_cross_section}
\end{align}
we can rewrite the reaction density in the form
\begin{align}
\gamma_{ij}(T)=\frac{T}{64\pi^4(1+\delta_{ij})}\int_{s_{ij}^{\min}}^\infty ds\,\sqrt{s}\,\widehat{\sigma}_{ij}(s)K_1\!\left(\frac{\sqrt{s}}{T}\right).
\label{eq:reduced_reaction_density}
\end{align}

The partial width for $h_2\to\chi\chi$ is given by
\begin{align}
\Gamma_{h_2\rightarrow\chi\chi}
&=\frac{y_\chi^2 M_{h_2}}{16\pi}\left(1-\frac{4M_\chi^2}{M_{h_2}^2}\right)^{3/2}
\Theta(M_{h_2}-2M_\chi),
\label{eq:DM_width}
\end{align}
where $\Theta$ is the Heaviside step function, and the symmetry factor for the two identical Majorana fermions is included. Assuming $h_2$ is in kinetic and chemical equilibrium with the bath, we can write its decay reaction density in the Maxwell–Boltzmann approximation as
\begin{align}
\gamma_{D,h_2}(T)
&=\int\frac{d^3p}{(2\pi)^3}e^{-E_2/T}\frac{M_{h_2}}{E_2}\Gamma_{h_2\rightarrow\chi\chi}\nonumber\\
&=\frac{M_{h_2}^2T}{2\pi^2}K_1\!\left(\frac{M_{h_2}}{T}\right) \Gamma_{h_2\rightarrow\chi\chi},\qquad E_2=\sqrt{\boldsymbol{p}^{\,2}+M_{h_2}^2}.
\label{eq:decay_reaction_density}
\end{align}
The factor $M_{h_2}/E_2$ accounts for time dilation, and the real scalar $h_2$ has one internal degree of freedom.

To avoid double counting on-shell $h_2$ exchange and explicit $h_2$ decays,
we subtract the real-intermediate-state (RIS) contribution from resonant scattering~\cite{Kolb:1979ui,Kolb:1979qa,Fry:1980ph}. In the narrow-width approximation, with $h_2$ in thermal equilibrium, this contribution is~\cite{Ala-Mattinen:2023rbm}
\begin{equation}
\gamma_{11}^{\mathrm{RIS}}(T)=\frac{M_{h_2}^2T}{2\pi^2} K_1\!\left(\frac{M_{h_2}}{T}\right)\frac{\Gamma_{h_2\to h_1h_1}\Gamma_{h_2\to\chi\chi}}{\Gamma_{h_2}}
=\operatorname{Br}(h_2\to h_1h_1)\gamma_{D,h_2}(T).
\end{equation}

Considering the dominant contributions including the ones induced by two-scalar mixing, the total DM particle-number production rate per unit volume can be written as
\begin{align}
R_\chi(T)=2\Bigl[
&\gamma_{11}^{\mathrm{full}}(T)+\gamma_{22}^{\mathrm{full}}(T)+\gamma_{12}^{\mathrm{full}}(T)+\gamma_{WW}^{\mathrm{full}}(T)+\gamma_{ZZ}^{\mathrm{full}}(T)+\gamma_{t\bar t}^{\mathrm{full}}(T)\nonumber\\
&-\gamma_{11}^{\mathrm{RIS}}(T)-\gamma_{WW}^{\mathrm{RIS}}(T)-\gamma_{ZZ}^{\mathrm{RIS}}(T)-\gamma_{t\bar t}^{\mathrm{RIS}}(T)+\gamma_{D,h_1}(T)+\gamma_{D,h_2}(T)\Bigr].
\label{total_rate}
\end{align}
Here, $\gamma_{ab}^{\mathrm{full}}$ denotes the full scattering reaction density before RIS subtraction, and the factor of two accounts for the two DM particles produced in each reaction. $\gamma_{D,h_i}$ ($i=1,\;2$) denotes the reaction density for $h_i\to\chi\chi$. The RIS contribution is given by $\gamma_{ab}^{\mathrm{RIS}}$, subtracting this contribution avoids double counting between resonant scattering and explicit mediator decays. For those contributions proportional to the two-scalar mixing, the concrete calculations of amplitudes, cross sections, decay widths, and thermal reaction densities are given in appendix~\ref{app:fullprod}.


\section{Cosmological evolution and relic abundance}
\label{sec:boltzmann}
Having obtained the DM production rate, we now follow the evolution of the DM abundance during reheating and the subsequent radiation-dominated epoch. Depending on the interplay between number-changing interactions, cosmological expansion, and entropy production, $\chi$ may be produced through FI or reach chemical equilibrium before undergoing UFO or WIMP-like FO. We use the Boltzmann equation to distinguish freeze-in from equilibrium histories and then classify the latter by their decoupling conditions.

\subsection{Boltzmann evolution during reheating}
\label{subsec:boltzmann_evolution}
We consider perturbative reheating during which the oscillating inflaton condensate has an approximately matter-like equation of state. During the inflaton-dominated cooling stage, the inflaton and radiation energy densities scale approximately as $\rho_\varphi\propto a^{-3}$ and $\rho_R\propto a^{-3/2}$, respectively~\cite{Garcia:2020wiy}. Taking the effective relativistic degrees of freedom to be constant, the radiation temperature therefore scales as $T\propto a^{-3/8}$. After reheating, the universe becomes radiation dominated, and conservation of comoving entropy yields $T\propto a^{-1}$.

We treat the effective number of relativistic degrees of freedom $g_*$ as constant and normalize the scale factor such that $a_{\mathrm{RH}}\equiv a(T_{\mathrm{RH}})=1$. The temperature evolution is then approximated by
\begin{align}
    T(a)=T_{\rm RH}
    \begin{cases}
        a^{-3/8}, & a<1,\\[2pt]
        a^{-1},   & a\geq1.
    \end{cases}
    \label{eq:temperature_scale_factor}
\end{align}
For the Hubble expansion rate, we adopt a piecewise approximation matched continuously at $T=T_{\mathrm{RH}}$~\cite{Kaneta:2019zgw},
\begin{align}
    H(T)=H_{\rm RH}
    \begin{cases}
        \displaystyle
        \left(\frac{T}{T_{\rm RH}}\right)^4,
        & T>T_{\rm RH},\\[6pt]
        \displaystyle
        \left(\frac{T}{T_{\rm RH}}\right)^2,
        & T\leq T_{\rm RH},
    \end{cases}
    \qquad
    H_{\rm RH}=\sqrt{\frac{\pi^2g_*}{90}}\,\frac{T_{\rm RH}^2}{M_{\rm P}},
    \label{eq:hubble_reheating}
\end{align}
where $M_{\mathrm{P}}\simeq2.4\times10^{18}~\mathrm{GeV}$ is the reduced Planck mass. The change in the temperature dependence of $H$ across reheating affects the competition between number-changing interactions and cosmic expansion, and hence the conditions for chemical equilibration and decoupling.

The DM number density evolves according to the Boltzmann equation~\cite{Cosme:2021baj}
\begin{align}
    \frac{dn_\chi}{dt}+3Hn_\chi=R_\chi(T)\left[
        1-\left(\frac{n_\chi}{n_\chi^{\rm eq}}\right)^2\right],
    \label{eq:boltzmann_nchi}
\end{align}
where $R_\chi(T)$ is the total DM particle-number production rate per unit volume defined in Eq.~\eqref{total_rate}. The quadratic term accounts for inverse reactions through detailed balance. Throughout the calculation, we adopt Maxwell--Boltzmann statistics~\cite{Chakraborty:2021azg}, for which the equilibrium DM number density is
\begin{align}
    n_\chi^{\rm eq}(T)
    =
    \frac{g_\chi m_\chi^2T}{2\pi^2}
    K_2\!\left(\frac{M_\chi}{T}\right),
    \qquad g_\chi=2.
    \label{eq:neq_chi}
\end{align}
with $g_\chi$ counting the internal degrees of freedom of the Majorana fermion. All particles except $\chi$ are assumed to remain in thermal equilibrium. To close the number-density equation, the inverse collision terms are evaluated with the kinetic-equilibrium ansatz $f_\chi(p,T)=(n_\chi/n_\chi^{\rm eq})e^{-E_\chi(p)/T}$. This ansatz retains the equilibrium momentum dependence at the bath temperature while allowing the normalization to depart from chemical equilibrium.

It is convenient to define the comoving DM number
\begin{align}
    Y_\chi\equiv n_\chi a^3,\;
    Y_\chi^{\rm eq}\equiv n_\chi^{\rm eq}a^3.
    \label{eq:comoving_number}
\end{align}
Using $\dot a=aH$, we rewrite Eq.~\eqref{eq:boltzmann_nchi} as
\begin{align}
    \frac{dY_\chi}{da}=\frac{a^2R_\chi(a)}{H(a)}\left[1-\left(\frac{Y_\chi}{Y_\chi^{\rm eq}}\right)^2\right].
    \label{eq:boltzmann_comoving}
\end{align}
We assume a vanishing initial DM abundance, $Y_\chi(a_i)=0$, at a temperature $T_i=T(a_i)>T_{\mathrm{RH}}$ on the cooling branch. Starting from this initial condition, we integrate Eq.~\eqref{eq:boltzmann_comoving} through reheating and into radiation domination until number-changing reactions no longer appreciably modify $Y_\chi$. The asymptotic comoving abundance is then evaluated as $Y_{\chi,\infty}\simeq Y_\chi(a_f)$, where $T_f=T(a_f)<T_{\mathrm{RH}}$.

\subsection{Identification of FI, UFO, and WIMP-like FO}
\label{subsec:production_regimes}
The primary distinction between the production regimes is whether DM attains chemical equilibrium with the thermal bath. A characteristic number-changing interaction rate evaluated at equilibrium can be defined as
\begin{equation}
    \Gamma_{\rm int}(T)
    \equiv
    \frac{R_\chi(T)}{n_\chi^{\rm eq}(T)}.
    \label{eq:interaction_rate}
\end{equation}
The ratio $\Gamma_{\mathrm{int}}/H$ measures the efficiency of these interactions relative to cosmological expansion. We determine the production regime from the solution of Eq.~\eqref{eq:boltzmann_comoving} by examining whether $Y_\chi$ approaches and subsequently tracks $Y_\chi^{\mathrm{eq}}$. For histories in which chemical equilibrium is established, we define $T_{\mathrm{FO}}$ as the temperature at the final departure from equilibrium and introduce $x_{\mathrm{FO}}=M_\chi/T_{\mathrm{FO}}$.

\paragraph{Freeze-in.}
DM is produced through FI if number-changing interactions do not establish chemical equilibrium with the thermal bath. In the dilute regime, where $Y_\chi\ll Y_\chi^{\mathrm{eq}}$ throughout the epoch of efficient production, inverse reactions are negligible. With the initial condition $Y_\chi(a_i)=0$, Eq.~\eqref{eq:boltzmann_comoving} then gives
\begin{align}
    Y_{\chi,\infty}^{\rm FI}\approx\int_{a_i}^{a_f}da\,\frac{a^2R_\chi(a)}{H(a)}.
    \label{eq:freezein_scale_factor}
\end{align}
Using Eq.~\eqref{eq:temperature_scale_factor}, the contributions accumulated during reheating and radiation domination can be written as
\begin{align}
    Y_{\chi,\infty}^{\rm FI}
    \approx\frac{8}{3}\int_{T_{\rm RH}}^{T_i}{\rm d}T
    \left(\frac{T_{\rm RH}}{T}\right)^8\frac{R_\chi(T)}{H(T)T}+
    \int_{T_f}^{T_{\rm RH}}{\rm d}T\left(\frac{T_{\rm RH}}{T}\right)^3\frac{R_\chi(T)}{H(T)T}.
    \label{eq:freezein_two_epochs}
\end{align}
The relative importance of these contributions depends on the temperature dependence of $R_\chi(T)$. Mediator decays, resonant scattering, and kinematic thresholds can concentrate production within different temperature intervals. We therefore retain the full temperature dependence of $R_\chi(T)$ in the numerical calculation, rather than impose a single power-law approximation over the entire evolution.

\paragraph{Ultra-relativistic freeze-out.}
If DM reaches chemical equilibrium, its subsequent evolution depends on when number-changing interactions can no longer maintain equilibrium. To illustrate the conditions for UFO, we first consider decoupling while DM is relativistic, $T_{\mathrm{FO}}>M_\chi$. Such decoupling during reheating requires interactions to become inefficient relative to expansion after equilibrium has been established. In a temperature interval where $R_\chi\propto T^p$ and $T\gg M_\chi$, the equilibrium number density scales as $n_\chi^{\mathrm{eq}}\propto T^3$, giving
\begin{align}
    \frac{\Gamma_{\rm int}}{H}\propto
    \begin{cases}
        T^{p-7}, & T>T_{\rm RH},\\
        T^{p-5}, & T<T_{\rm RH}.
    \end{cases}
    \label{eq:relativistic_decoupling_scaling}
\end{align}
Thus, during reheating with a matter-like inflaton equation of state, a local scaling with $p>7$ causes $\Gamma_{\mathrm{int}}/H$ to decrease as the bath cools and can allow relativistic decoupling. In the scalar-mediated model, whether this occurs must be determined from the total scattering and decay contributions to $R_\chi(T)$, including their threshold dependence.

The evolution after chemical decoupling differs from that of a relativistic relic in radiation domination. While DM remains relativistic during reheating, its equilibrium comoving abundance grows as
\begin{align}
    Y_\chi^{\rm eq}\propto a^3T^3\propto a^{15/8}.
    \label{eq:relativistic_equilibrium_growth}
\end{align}
After decoupling, the DM abundance can fall below this increasing equilibrium value. Residual interactions can then continue to produce DM without restoring chemical equilibrium. If inverse reactions provide only a small correction to the integrated production after decoupling, the final comoving abundance can be estimated as
\begin{align}
    Y_{\chi,\infty}^{\rm UFO}\simeq Y_{\rm FO}+\int_{a_{\rm FO}}^{a_f}
    da\,\frac{a^2R_\chi[T(a)]}{H[T(a)]},
    \label{eq:ufo_abundance_estimate}
\end{align}
where $Y_{\rm FO}\equiv Y_\chi(a_{\rm FO}) \simeq Y_\chi^{\rm eq}(a_{\rm FO})$ is the comoving DM abundance at chemical decoupling.

Eqs.~\eqref{eq:freezein_scale_factor} and~\eqref{eq:ufo_abundance_estimate} exhibit the connection between FI and UFO. Both mechanisms can involve an extended period of out-of-equilibrium production, but UFO inherits a nonzero abundance from an earlier equilibrium phase. Depending on the temperature dependence of the production rate, continued production after decoupling can contribute substantially to, or even dominate, the final abundance. In such cases, the final abundance cannot be inferred solely from the equilibrium abundance at $T_{\mathrm{FO}}$.

\paragraph{WIMP-like freeze-out.}
WIMP-like FO occurs when chemical equilibrium persists into the nonrelativistic regime. The equilibrium number density then becomes Boltzmann suppressed
\begin{align}
    n_\chi^{\rm eq}(T)\simeq  g_\chi\left(\frac{m_\chi T}{2\pi}\right)^{3/2}e^{-m_\chi/T}.
    \label{eq:nonrelativistic_equilibrium}
\end{align}
When annihilation becomes too slow to maintain chemical equilibrium, $Y_\chi$ departs from and exceeds $Y_\chi^{\mathrm{eq}}$. Residual annihilation continue to reduce the comoving abundance after this departure. Once $Y_\chi\gg Y_\chi^{\mathrm{eq}}$, the Boltzmann equation reduces to
\begin{align}
    \frac{d}{da}\left(\frac{1}{Y_\chi}\right)\simeq\frac{R_\chi(a)}{a^4H(a)
    \left[n_\chi^{\rm eq}(a)\right]^2}.
    \label{eq:residual_annihilation}
\end{align}
The final comoving abundance therefore depends on both the abundance at departure from equilibrium and the subsequent annihilation history. If the freeze out of DM occurs during reheating, continued entropy production further reduces the entropy-normalized abundance.

For decoupling during reheating, the maximum of the equilibrium comoving abundance provides an approximate boundary between UFO and WIMP-like FO. Defining $x=M_\chi/T$, we obtain
\begin{align}
    Y_\chi^{\rm eq}\propto x^7K_2(x)\simeq\sqrt{\frac{\pi}{2}}x^{13/2}e^{-x},
    \qquad x\gg1.
    \label{eq:equilibrium_turnover}
\end{align}
The nonrelativistic approximation peaks at $x\simeq6.24$ for the full Maxwell--Boltzmann expression. Following Ref.~\cite{Henrich:2025pca}, we adopt $x=13/2$ as an operational classification criterion: decoupling during reheating with $x_{\mathrm{FO}}<13/2$ is classified as UFO, whereas $x_{\mathrm{FO}}\geq13/2$ is classified as WIMP-like FO. 

For chemical decoupling at or after the completion of reheating, $T_{\mathrm{FO}}\leq T_{\mathrm{RH}}$, the equilibrium comoving abundance instead scales as $Y_\chi^{\mathrm{eq}}\propto x^2K_2(x)$. It approaches a constant as $x\to0$ and decreases monotonically with increasing $x>0$. There is therefore no turnover at finite $x$, and the classification criterion based on the maximum of $Y_\chi^{\mathrm{eq}}$ during reheating does not apply. We instead adopt the operational classification
\begin{align}
    \begin{cases}
        x_{\rm FO}\leq1,
        & \text{UFO},\\
        x_{\rm FO}>1,
        & \text{WIMP-like FO}.
    \end{cases}
    \label{eq:RD_freezeout_classification}
\end{align}
Under this convention, semi-relativistic FO with $1<x_{\mathrm{FO}}\leq3$ is included in the WIMP-like FO category. The UFO label encompasses relativistic decoupling, while the strictly ultra-relativistic limit corresponds to $x_{\mathrm{FO}}\ll1$.

The analytical expressions above describe the limiting behavior of the different production regimes. In the numerical analysis, we solve the full Boltzmann equation, Eq.~\eqref{eq:boltzmann_comoving}, to resolve the transition regions between these regimes and account for continued DM production or annihilation after chemical decoupling.

\subsection{Present-day relic abundance}
\label{subsec:present_relic_abundance}
After reheating, the comoving entropy of the thermal bath is conserved. Once DM number-changing reactions also become ineffective, the entropy-normalized abundance approaches a constant value,
\begin{align}
    \mathcal{Y}_{\chi,\infty}\equiv\frac{n_\chi(T_f)}{s(T_f)},
    \qquad s(T)=\frac{2\pi^2}{45}g_{*s}T^3,
    \label{eq:entropy_normalized_abundance}
\end{align}
where $g_{*s}$ denotes the effective number of relativistic degrees of freedom contributing to the entropy density. Throughout the calculation, we take $g_{*s}$ and $g_*$ to be constant. With the normalization $a_{\mathrm{RH}}=1$, entropy conservation implies $s(T)a^3(T)=s(T_{\mathrm{RH}})$ after reheating. The final entropy-normalized abundance is therefore related to the asymptotic comoving DM number density by
\begin{align}
    \mathcal{Y}_{\chi,\infty}=\frac{Y_{\chi,\infty}}{s(T_{\mathrm{RH}})}.
    \label{eq:comoving_to_entropy_yield}
\end{align}

The effect of entropy production is particularly transparent in the instantaneous-decoupling approximation. If FO occurs at $T_{\mathrm{FO}}>T_{\mathrm{RH}}$ and subsequent number-changing reactions are neglected, the final entropy-normalized abundance is approximately
\begin{align}
    \mathcal Y_{\chi,\infty}^{\rm inst}\simeq\frac{n_\chi^{\rm eq}(T_{\rm FO})}{s(T_{\rm FO})}\left(\frac{T_{\rm RH}}{T_{\rm FO}}\right)^5.
    \label{eq:reheating_entropy_dilution}
\end{align}
The dilution factor follows from $s(T)a^3(T)\propto T^{-5}$ during reheating and accounts for the entropy injected between chemical decoupling and the completion of reheating. In the full calculation, continued production and residual annihilation can respectively increase and decrease the comoving DM abundance. Hence their effects are included together with entropy dilution when determining the final entropy-normalized abundance.

The present-day DM relic abundance is given by
\begin{align}
    \Omega_\chi h^2&=\frac{M_\chi s_0}{\rho_{\mathrm{c}}/h^2}\simeq2.742\times10^8\left(\frac{M_\chi}{\mathrm{GeV}}\right)\mathcal{Y}_{\chi,\infty},
    \label{eq:relic_density}
\end{align}
where $s_0$ and $\rho_{\mathrm{c}}$ are the present-day entropy density and critical density, respectively. Using Eq.~\eqref{eq:comoving_to_entropy_yield}, we can equivalently express the relic abundance in terms of the comoving DM number density as
\begin{equation}
    \Omega_\chi h^2\simeq2.742\times10^8\left(\frac{M_\chi}{\mathrm{GeV}}\right)\frac{Y_{\chi,\infty}}{(2\pi^2/45)g_{*s}T_{\mathrm{RH}}^3}.
    \label{eq:relic_density_comoving}
\end{equation}
All values of $\Omega_\chi h^2$ presented below are obtained by solving the full Boltzmann equation and evolving the system until reheating has ended and the comoving DM abundance has stabilized.


\section{Numerical results}
\label{sec:numerical}
In this section, we determine the thermal histories and relic abundances for representative scalar-mediator parameters. Unless otherwise specified, we use
\begin{align}
M_{h_1}=125~\mathrm{GeV},\quad
\kappa=\kappa_{H\phi}=100~\mathrm{GeV},\quad\lambda_{H\phi}=10^{-5}
\label{eq:numerical_benchmark}
\end{align}
in the following computations. In models with a nonzero singlet VEV, the quartic portal coupling $\lambda_{H\phi}$ contributes to doublet--singlet mixing. Small but nonzero portal couplings are additionally motivated in classically scale-invariant constructions, which provide a framework for addressing the origin and hierarchy of the electroweak scale~\cite{Coleman:1973jx,Englert:2013gz,Khoze:2013oga}. Motivated by these broader considerations, we adopt $\lambda_{H\phi}=10^{-5}$ as a representative benchmark choice. 

The reaction densities are evaluated retaining their full mass dependence. All relic abundances are obtained by solving the full Boltzmann equation through reheating and the subsequent radiation-dominated epoch. We identify the production regimes from the numerical evolution using an operational criterion for sustained tracking of the equilibrium abundance. Specifically, we require
\begin{align}
\left|\frac{n_\chi}{n_\chi^{\mathrm{eq}}}-1\right|<0.1,
\quad\frac{\Gamma_{\mathrm{int}}}{H}>1,
\label{eq:numerical_equilibrium_criterion}
\end{align}
where $\Gamma_{\mathrm{int}}$ is defined in Eq.~\eqref{eq:interaction_rate}. To identify the sustained chemical-equilibrium phase required for both UFO and WIMP-like FO, both conditions must hold at every sampled point within a contiguous interval spanning $\Delta\ln a\geq0.2$. The abundance criterion permits deviations of less than $10\%$ from the instantaneous equilibrium density. The minimum duration excludes isolated crossings or brief passages through $n_\chi/n_\chi^{\mathrm{eq}}\simeq1$. The interaction-rate criterion provides an additional check that number-changing reactions are efficient relative to cosmological expansion.

Histories without a qualifying equilibrium interval are classified as FI. Otherwise, we define $a_{\mathrm{FO}}$ as the scale factor at the last sampled point of the final qualifying interval. The corresponding $T_{\mathrm{FO}}=T(a_{\mathrm{FO}})$ and $x_{\mathrm{FO}}=M_\chi/T_{\mathrm{FO}}$ are then used to distinguish UFO from WIMP-like FO according to the criteria in Sec.~\ref{subsec:production_regimes}.

\subsection{The production channels and representative thermal history}
\label{subsec:numerical_histories}
\begin{figure*}[t!]
\centering
    \includegraphics[width=0.49\textwidth]{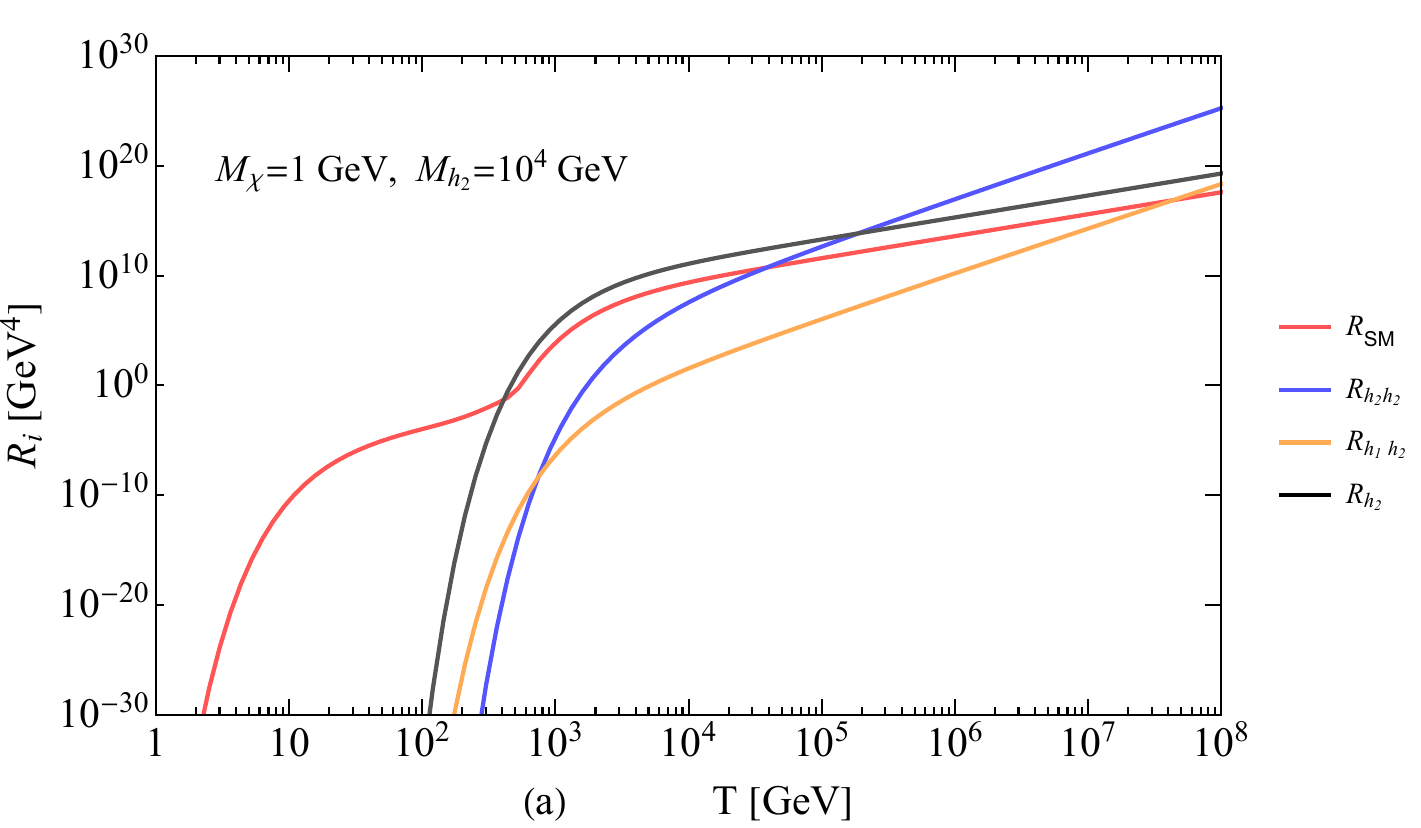}
    \includegraphics[width=0.49\textwidth]{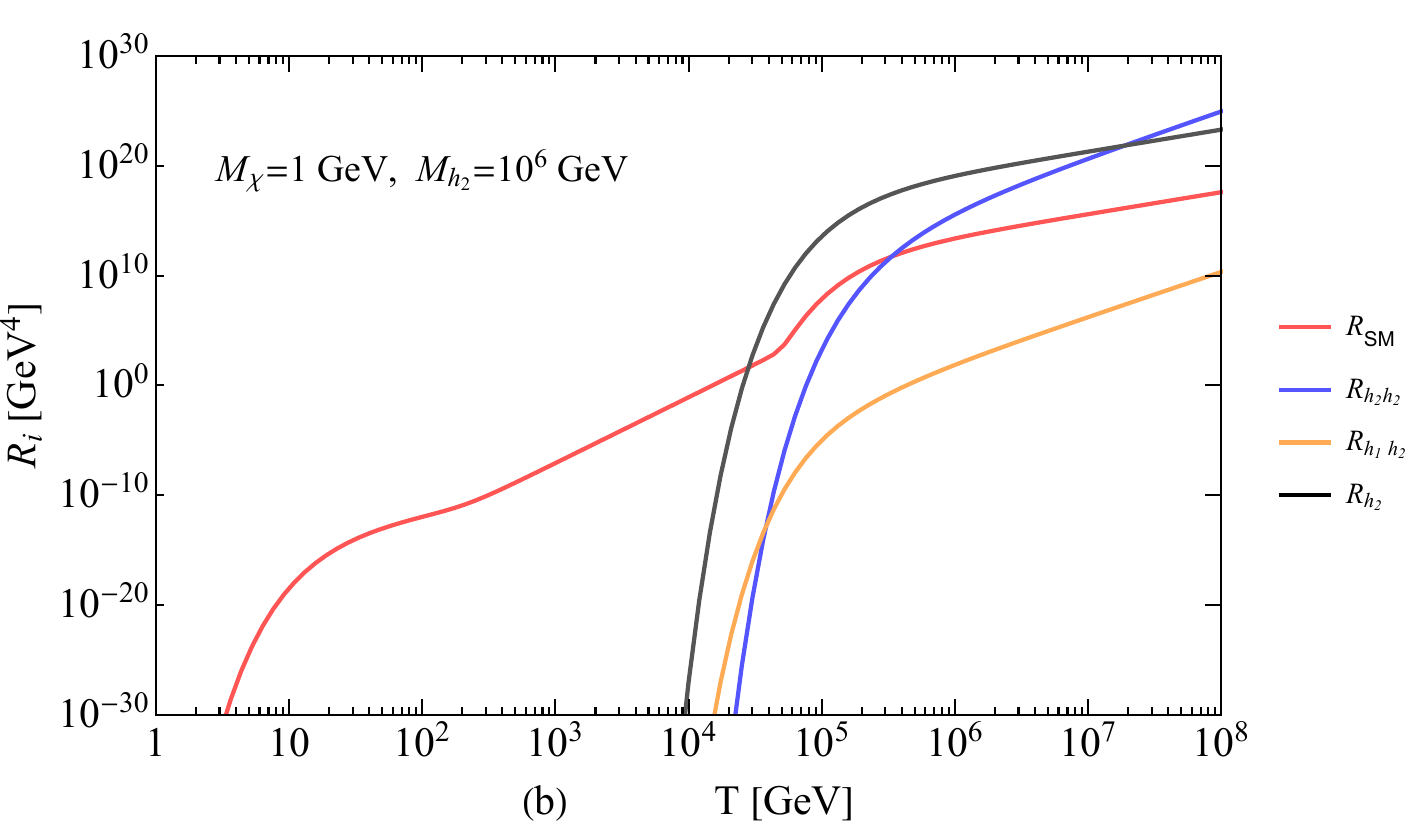}
    \includegraphics[width=0.49\textwidth]{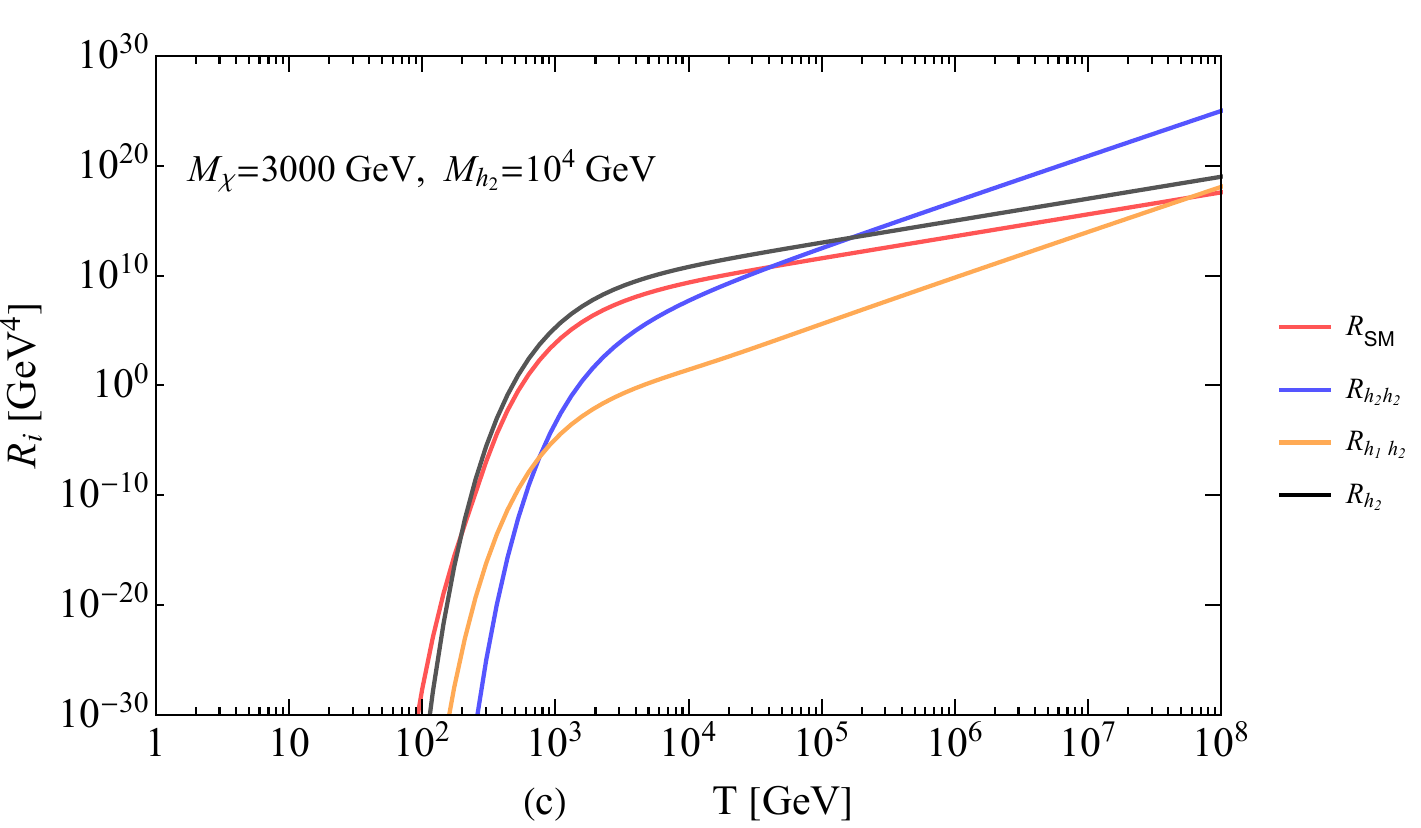}
    \includegraphics[width=0.49\textwidth]{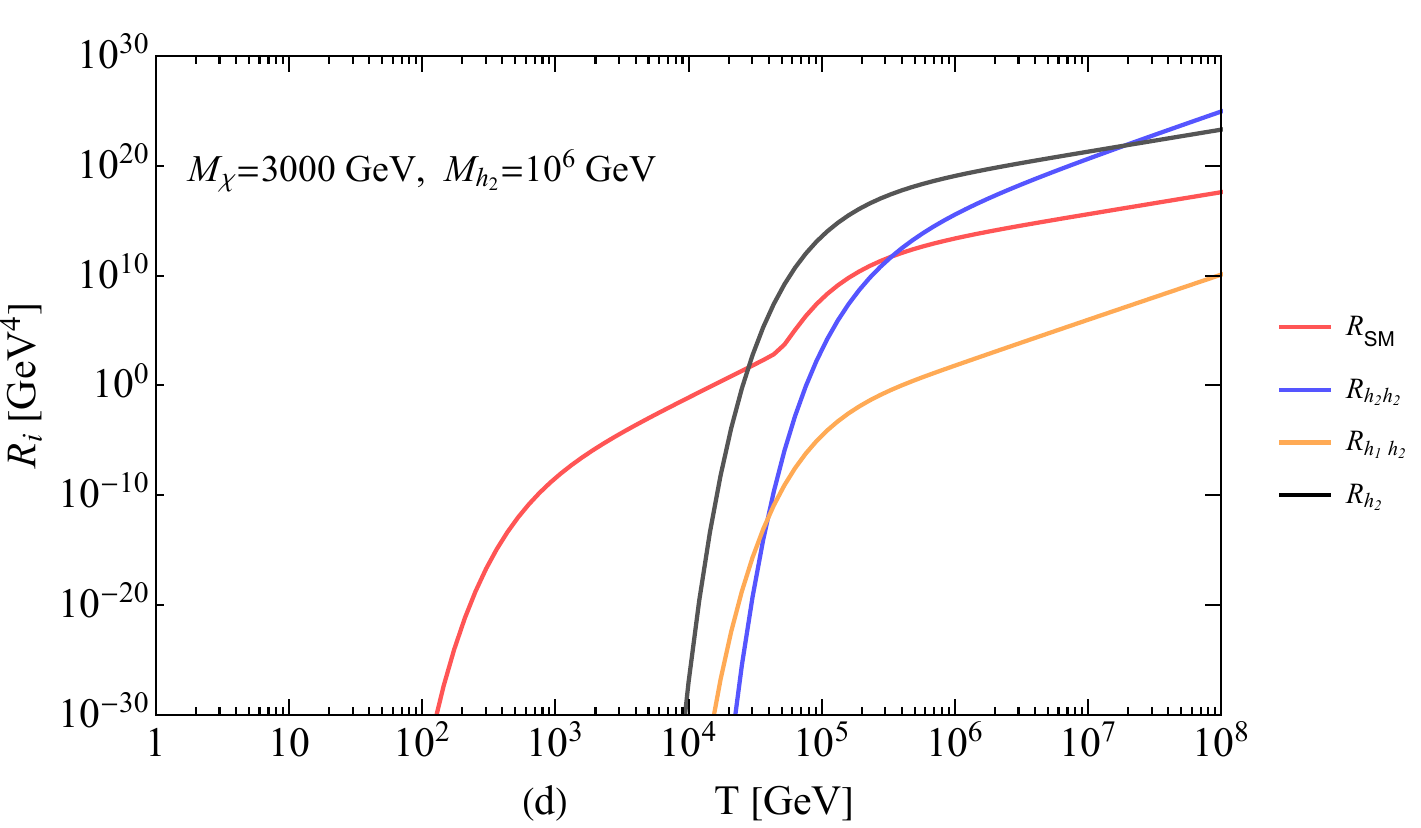}
    \caption{Production-rate contributions as functions of temperature for $y_\chi=0.1$. The red curves show $R_{\rm SM}$, comprising $h_1h_1$, $W^+W^-$, $ZZ$, $t\bar t$ annihilation, and $h_1\to\chi\chi$ when kinematically allowed. The blue, orange, and black curves show the contributions from $h_2h_2\to\chi\chi$, $h_1h_2\to\chi\chi$, and $h_2\to\chi\chi$, respectively. The mass choices are $(M_\chi,M_{h_2})=(1,10^4)$ (a), $(1,10^6)$ (b), $(3000,10^4)$ (c), and $(3000,10^6)~\mathrm{GeV}$ (d).}
\label{fig:production_rates}
\end{figure*}

To display the contributions from the different production channels, we define
\begin{align}
&R_{\rm SM}(T)=2\left[\gamma_{11}^{\rm full}(T)+\gamma_{WW}^{\rm full}(T)+\gamma_{ZZ}^{\rm full}(T)+\gamma_{t\bar t}^{\rm full}(T)+\gamma_{D,h_1}(T)\right],\nonumber\\
&R_{h_2h_2}(T)=2\gamma_{22}^{\rm full}(T),\;
R_{h_1h_2}(T)=2\gamma_{12}^{\rm full}(T),\;
R_{h_2}(T)=2\gamma_{D,h_2}(T),
\end{align}
and plot Fig.~\ref{fig:production_rates}. The scattering contributions shown here are evaluated before RIS subtraction. The total production rate entering the Boltzmann equation includes the corresponding subtraction terms. At sufficiently high temperatures, $h_2h_2\to\chi\chi$ provides the dominant contribution for the benchmarks shown. When both $h_2$ and $\chi$ are relativistic, the contribution from fermion exchange in the $t$ and $u$ channels scales parametrically as $R_{h_2h_2}\propto y_\chi^4T^4$, up to logarithmic corrections. By comparison, the contribution from the kinematically allowed decay $h_2\to\chi\chi$ scales as $R_{h_2}\propto \Gamma_{h_2\to\chi\chi}M_{h_2}T^2$ in the relativistic mediator limit, $T\gg M_{h_2}$. The faster growth of the scattering contribution with temperature therefore explains its dominance at sufficiently high temperatures.

As the temperature falls below $M_{h_2}$, the equilibrium number density of the heavy scalar becomes Boltzmann suppressed. The contribution from $h_2h_2\to\chi\chi$ decreases more rapidly than that from $h_2\to\chi\chi$, because the scattering process requires two heavy particles in the initial state, whereas the decay process $h_2\to\chi\chi$ involves only one. For the considered benchmarks, decays consequently become the largest individual source of DM over an intermediate temperature interval. The mixed channel $h_1h_2\to\chi\chi$ remains subleading for the benchmarks shown.

At lower temperatures, production from the SM bath can remain effective after the channels involving thermal $h_2$ particles have become suppressed. For example, in the regime $M_{h_1},M_\chi\ll T\ll M_{h_2}$, the $h_1h_1\to\chi\chi$ contribution scales as
\begin{align}
R_{h_1h_1}\propto\frac{y_\chi^2\kappa_{H\phi}^2}{M_{h_2}^4}T^6.
\label{eq:h1h1_rate_scaling}
\end{align}
Although this scaling does not describe the entire $R_{\rm SM}$ contribution,
it illustrates that the production from the SM bath at lower temperatures accounts for the extended low-temperature tails in Fig.~\ref{fig:production_rates}. For $M_\chi=1~\mathrm{GeV}$, the decay $h_1\to\chi\chi$ is open and dominates the sufficiently low-temperature part of the red curves in Figs.~\ref{fig:production_rates} (a) and (b). Its eventual suppression follows the depletion of the thermal $h_1$ population. For $M_\chi=3000~\mathrm{GeV}$, this decay is closed, and the low-temperature contribution instead arises from scattering. The larger DM pair-production threshold suppresses this contribution at higher temperatures, as seen in Figs.~\ref{fig:production_rates} (c) and (d).

The changing contributions of the different production channels are relevant to UFO because chemical equilibration and decoupling depend on the temperature evolution of the total interaction efficiency, $\Gamma_{\mathrm{int}}/H=R_\chi/(n_\chi^{\mathrm{eq}}H)$. A large production rate at high temperatures does not by itself guarantee chemical equilibrium, particularly during reheating, when $H\propto T^4$. Once equilibrium has been established, the Boltzmann suppression of the heavy-scalar population can rapidly reduce the contributions from channels involving thermal $h_2$ particles and lower $\Gamma_{\mathrm{int}}/H$. If this loss of interaction efficiency occurs while DM is still relativistic and the remaining SM-bath contributions are insufficient to sustain chemical equilibrium, DM can undergo UFO.

\begin{figure*}[t!]
\centering
    \includegraphics[width=0.49\textwidth]{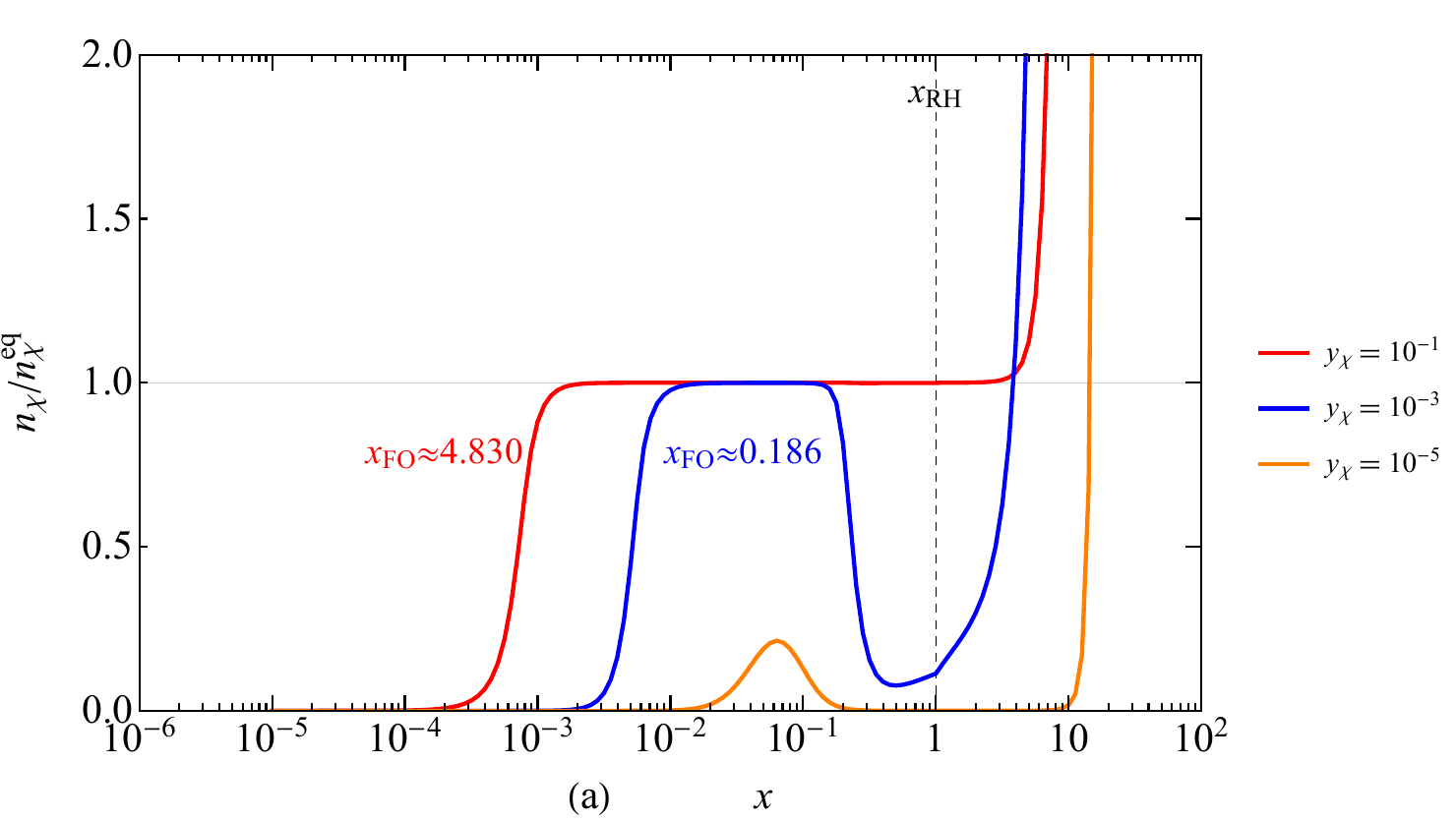}
    \includegraphics[width=0.49\textwidth]{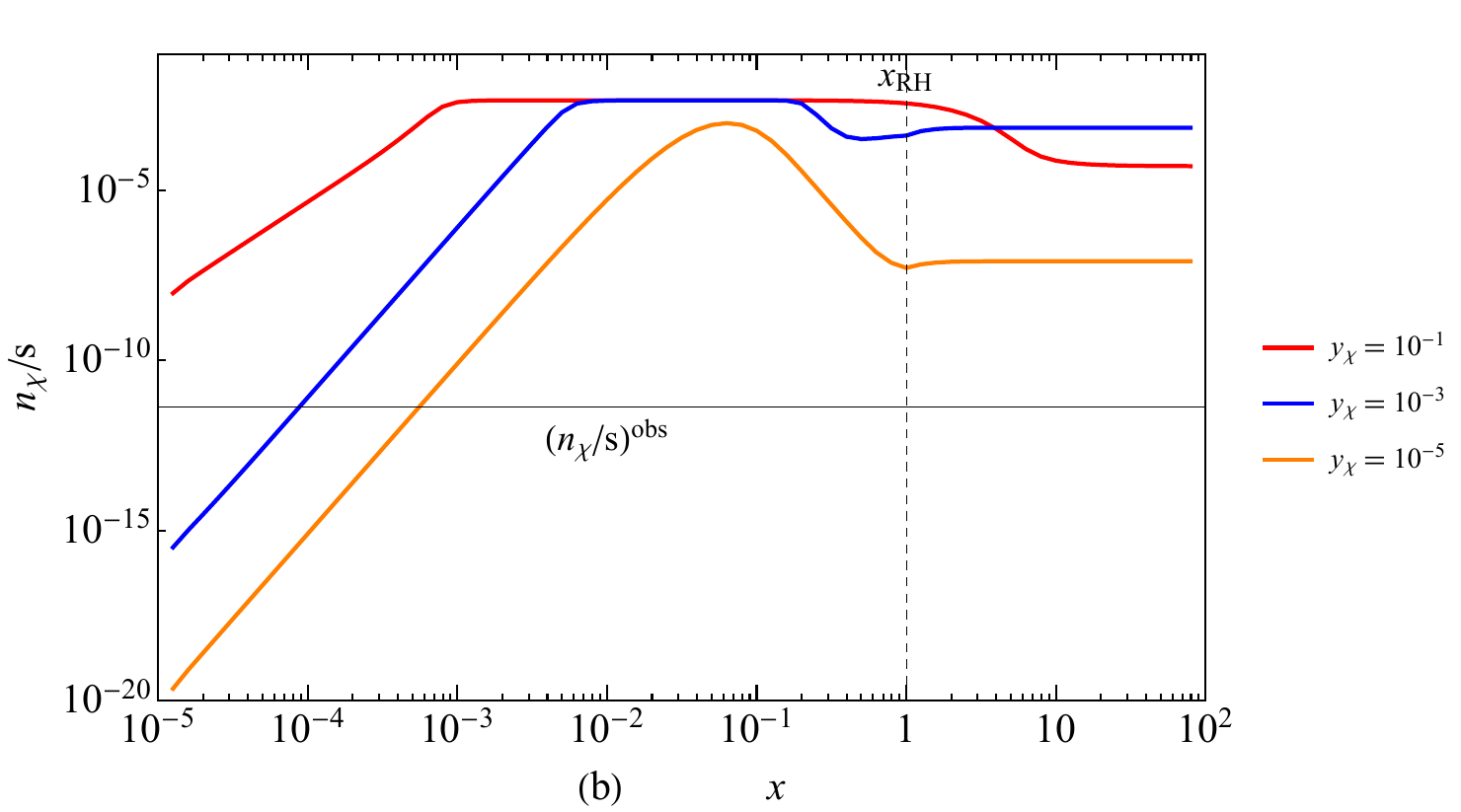}
    \includegraphics[width=0.49\textwidth]{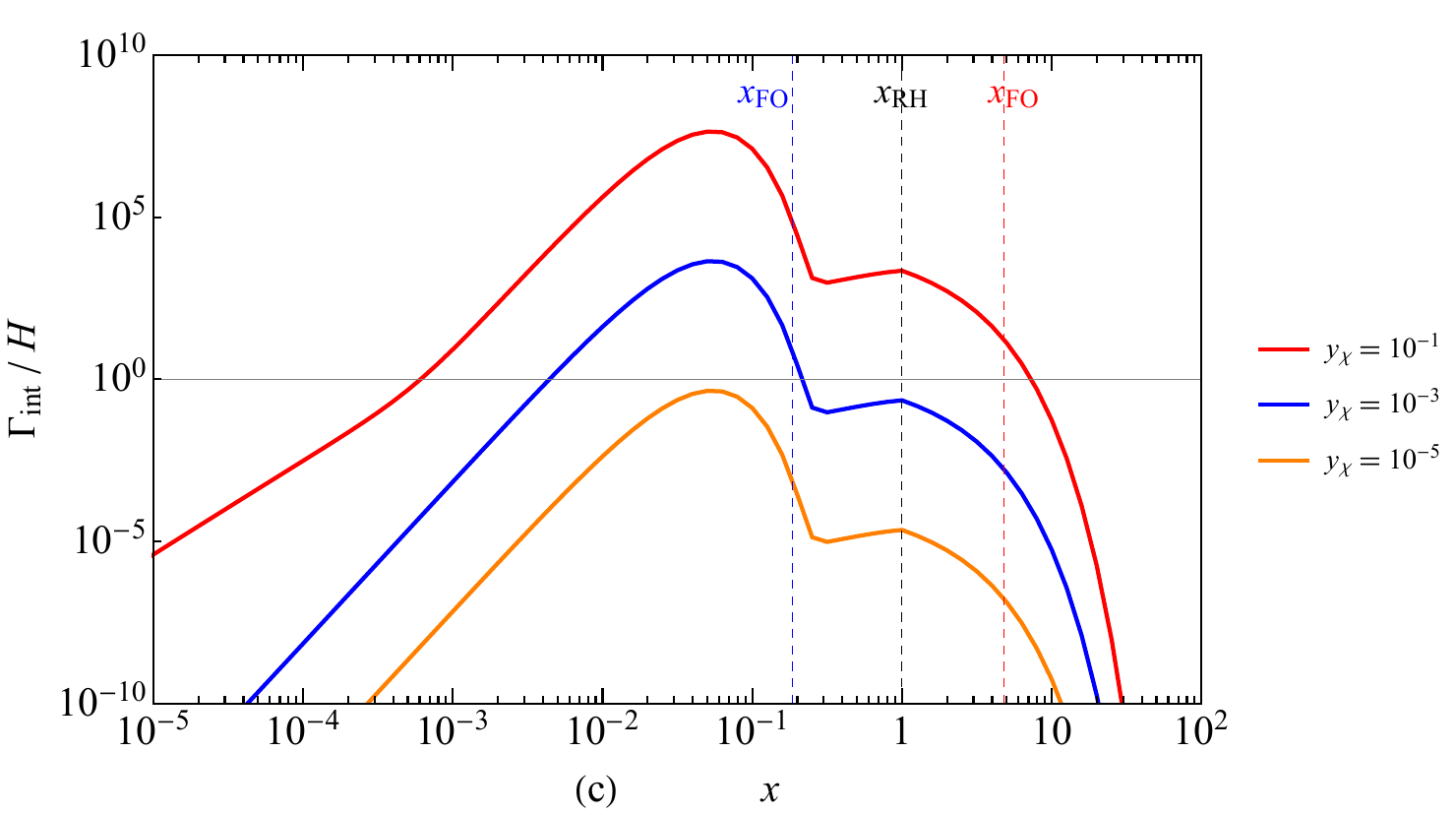}
    \includegraphics[width=0.49\textwidth]{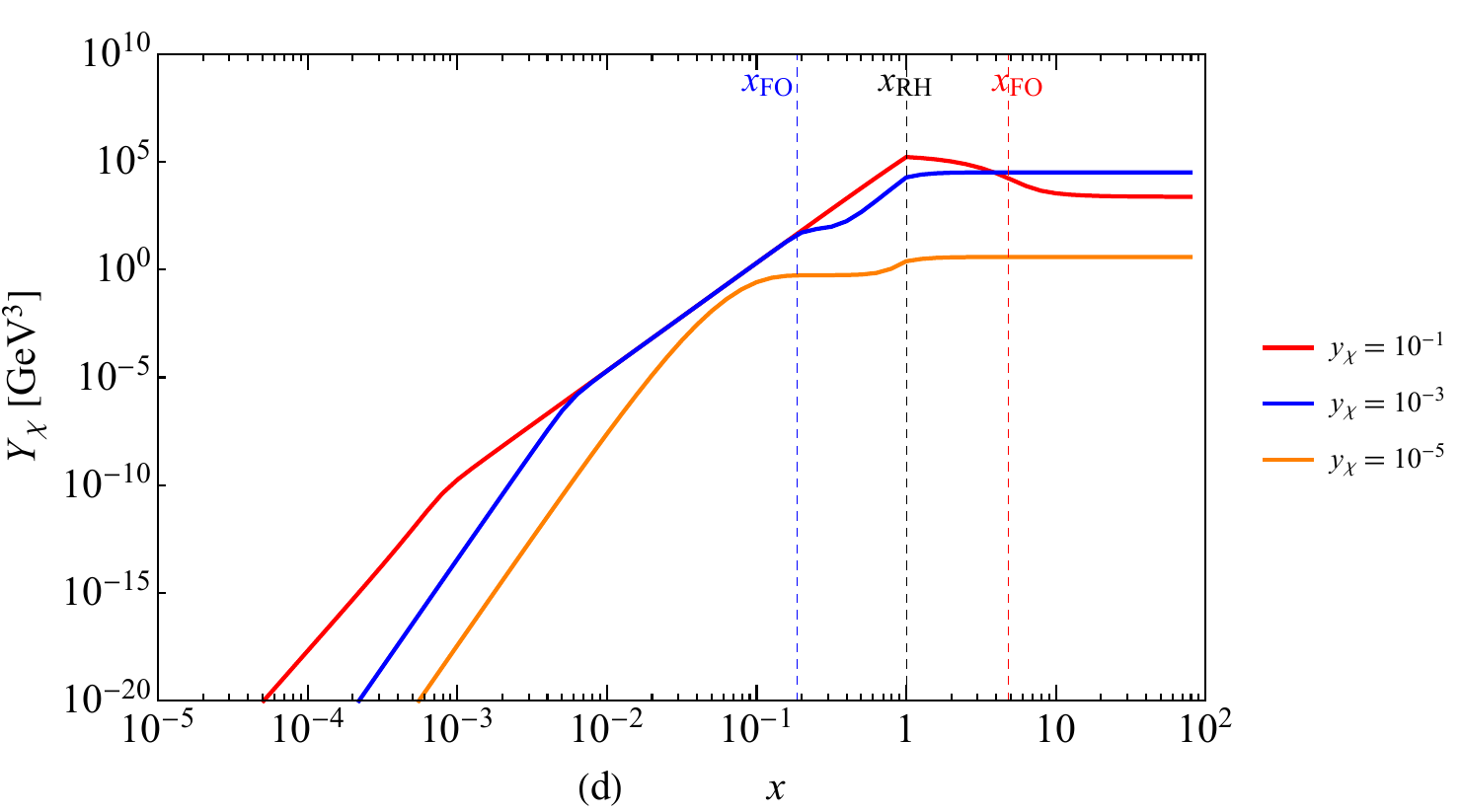}
    \caption{Representative DM histories for $M_\chi=T_{\mathrm{RH}}=100~\mathrm{GeV}$ and $M_{h_2}=10^4~\mathrm{GeV}$. The panels show (a) $n_\chi/n_\chi^{\mathrm{eq}}$, (b) the entropy-normalized abundance $n_\chi/s$, (c) the interaction efficiency $\Gamma_{\mathrm{int}}/H=R_\chi/(Hn_\chi^{\mathrm{eq}})$, and (d) the comoving number $Y_\chi=n_\chi a^3$, as functions of $x=M_\chi/T$. The red, blue, and orange curves correspond to $y_\chi=10^{-1}$, $10^{-3}$, and $10^{-5}$, respectively. The black, blue and red vertical dashed lines mark $x_{\mathrm{RH}}=1$, $x_{\mathrm{FO}}=0.186$ and $x_{\mathrm{FO}}=4.83$, respectively. The horizontal reference lines in (a), (b), and (c) denote chemical equilibrium, the yield corresponding to $\Omega_\chi h^2=0.12$, and $\Gamma_{\mathrm{int}}/H=1$, respectively.}
\label{fig:thermal_histories}
\end{figure*}

To illustrate this connection clearly, we consider the benchmark
\begin{align}
M_\chi=100~\mathrm{GeV},\quad T_{\mathrm{RH}}=100~\mathrm{GeV},\quad M_{h_2}=10^4~\mathrm{GeV},
\label{eq:thermal_history_benchmark}
\end{align}
and plot Fig.~\ref{fig:thermal_histories} with $y_\chi=10^{-1}$ (red curves), $10^{-3}$ (blue curves), and $10^{-5}$ (orange curves). Figs.~\ref{fig:thermal_histories} (a) and (c) show $n_\chi/n_\chi^{\mathrm{eq}}$ and $\Gamma_{\mathrm{int}}/H$, respectively, which are used to identify whether chemical equilibrium is established and maintained. Figs.~\ref{fig:thermal_histories} (b) and (d) display the evolution of the entropy-normalized abundance $n_\chi/s$ and the comoving abundance $Y_\chi$, respectively. For this benchmark, reheating ends at $x_{\mathrm{RH}}\equiv M_\chi/T_{\mathrm{RH}}=1$. And the line $\Gamma_{\mathrm{int}}/H=1$ provides a reference for comparing interaction and expansion time scales.

The intermediate-coupling benchmark, $y_\chi=10^{-3}$ (blue curves), illustrates UFO. As the bath cools, the interaction efficiency shown in Fig.~\ref{fig:thermal_histories} (c) becomes sufficient for DM to establish chemical equilibrium, as indicated by the interval with $n_\chi/n_\chi^{\mathrm{eq}}\simeq1$ in Fig.~\ref{fig:thermal_histories} (a). Further cooling suppresses the contributions from channels involving thermal heavy scalars, reducing the interaction efficiency and leading to decoupling at $x_{\mathrm{FO}}\simeq0.186$. This corresponds to $T_{\mathrm{FO}}\simeq5.4\times10^2~\mathrm{GeV}$, so DM decouples during reheating while still relativistic. The marked decoupling point is identified using the sustained-equilibrium criteria specified above.

The evolution after decoupling in Fig.~\ref{fig:thermal_histories} (d) is essential for determining the final UFO abundance. The blue curve shows that the comoving DM number density $Y_\chi=n_\chi a^3$ continues to increase substantially after chemical decoupling, even though it no longer tracks $Y_\chi^{\mathrm{eq}}$. Approximating $Y_{\chi,\infty}$ by its value at decoupling would therefore underestimate the final comoving abundance for UFO. Meanwhile, Fig.~\ref{fig:thermal_histories} (b) shows that the entropy-normalized abundance $\mathcal{Y}_\chi=n_\chi/s$ decreases over part of the same interval. These behaviors are consistent because
\begin{align}
    \mathcal{Y}_\chi=\frac{Y_\chi}{sa^3}, \quad sa^3\propto a^{15/8}
    \qquad\text{during reheating}.
    \label{eq:yield_entropy_relation}
\end{align}
When the comoving entropy grows faster than $Y_\chi$, the entropy-normalized abundance decreases despite continued net DM production. Fig.~\ref{fig:thermal_histories} (d) therefore directly displays production after decoupling, while Fig.~\ref{fig:thermal_histories} (b) shows its competition with entropy dilution during reheating. In addition, the final relic abundance of the UFO benchmark exceeds that of the larger-coupling WIMP-like FO benchmark. In the latter case, interactions maintain equilibrium to a lower temperature and allow more efficient depletion through annihilation. In the UFO history, continued production after decoupling instead increases the comoving abundance. The final abundance therefore need not increase monotonically with the Yukawa coupling.

The other two benchmarks illustrate WIMP-like FO and FI. For $y_\chi=10^{-1}$, interactions maintain chemical equilibrium beyond the end of reheating, with the final departure occurring at $x_{\rm FO}\simeq4.83$, corresponding to $T_{\rm FO}\simeq20.7~\mathrm{GeV}$. This history is therefore classified as WIMP-like FO. As freeze-out approaches, the comoving abundance decreases while tracking the falling equilibrium value, and residual annihilation continues to deplete it after chemical decoupling. For $y_\chi=10^{-5}$, $\Gamma_{\mathrm{int}}/H$ remains below unity, as shown by the orange curve in Fig.~\ref{fig:thermal_histories} (c). Chemical equilibrium is not established during production, and this benchmark realizes FI. The late-time increase in $n_\chi/n_\chi^{\mathrm{eq}}$ results from the exponential decrease of the equilibrium density and does not indicate chemical equilibration.

These benchmarks are chosen to illustrate the thermal histories of FI, WIMP-like FO, and UFO, rather than to reproduce the observed relic abundance. Their final entropy-normalized abundances all exceed the reference value $\mathcal{Y}_{\chi,\mathrm{obs}}\simeq4.38\times10^{-12}$ for $M_\chi=100~\mathrm{GeV}$. Parameter choices yielding $\Omega_\chi h^2=0.12$ are examined in the next subsection.

\subsection{Relic abundance and production regimes}
\label{subsec:numerical_parameter_space}

\begin{figure*}[t!]
\centering
    \includegraphics[width=0.49\textwidth]{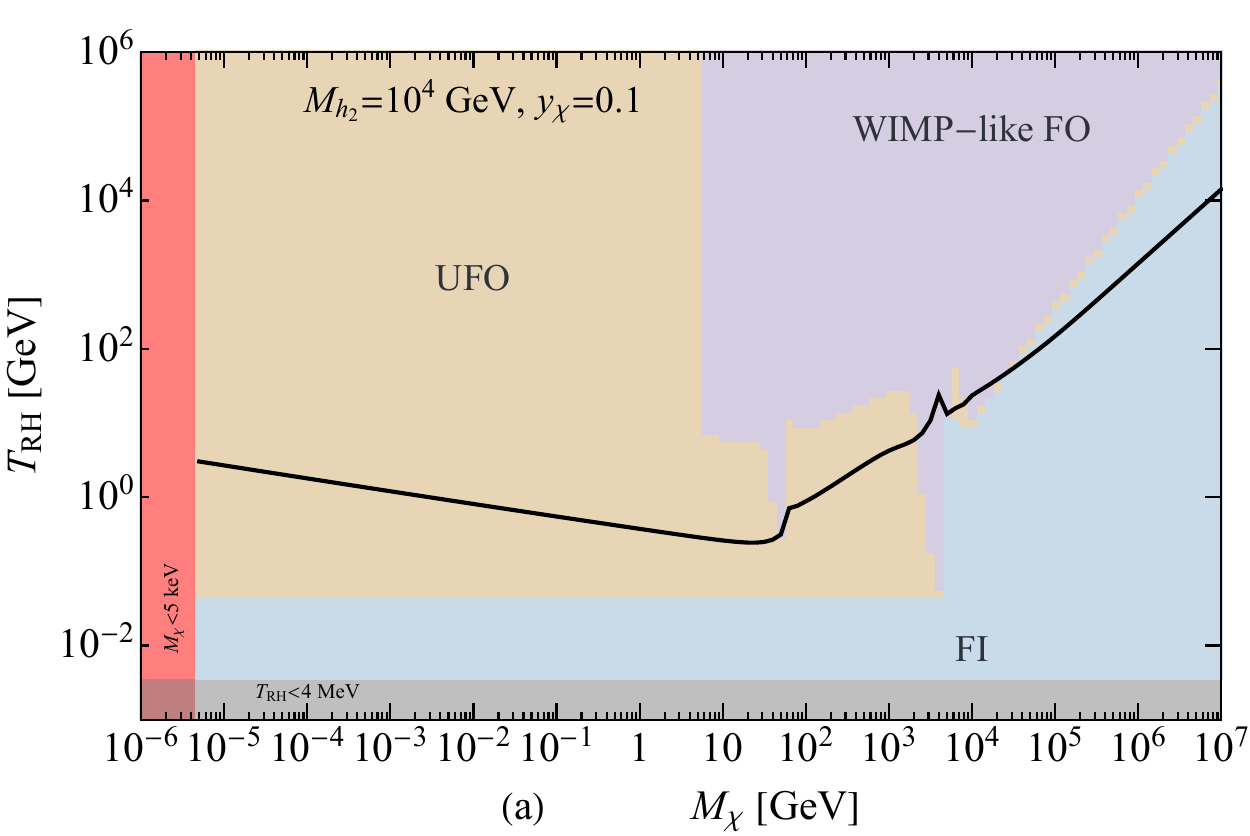}
    \includegraphics[width=0.49\textwidth]{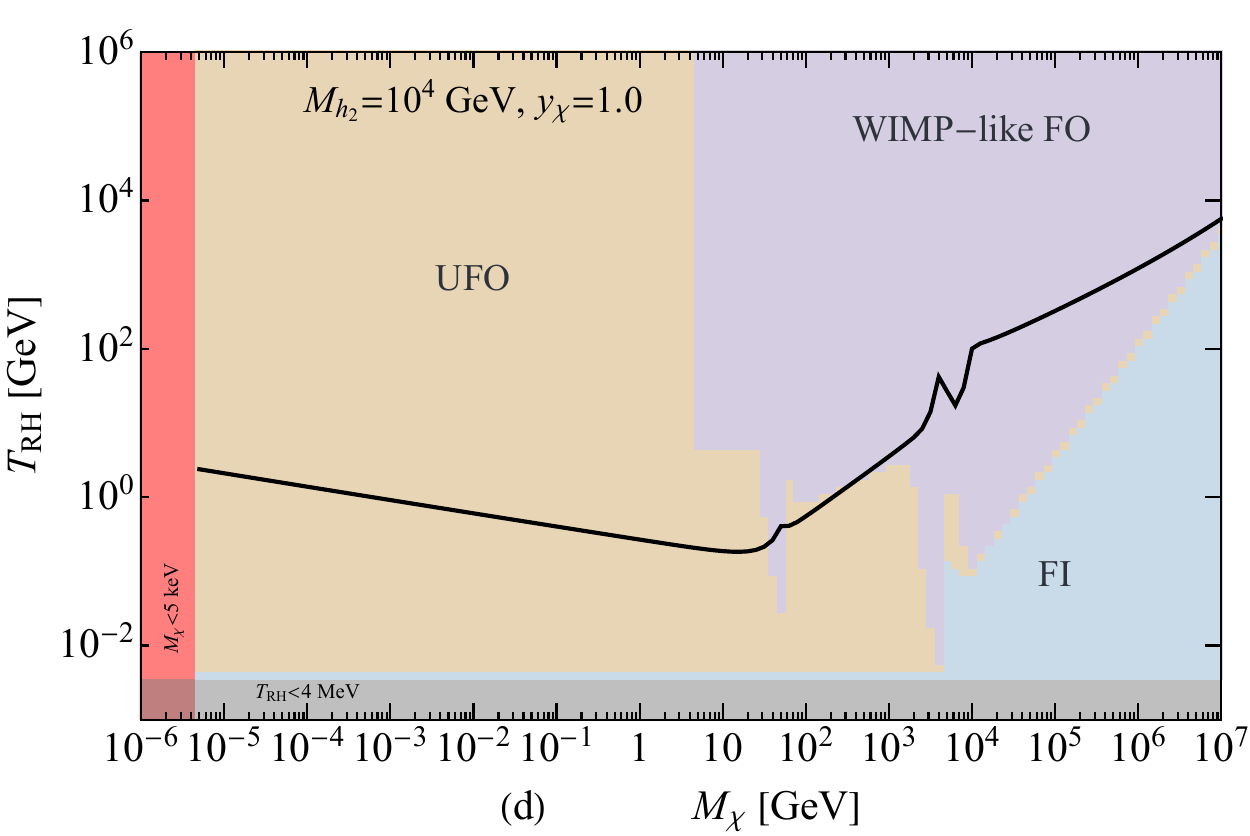}
    \includegraphics[width=0.49\textwidth]{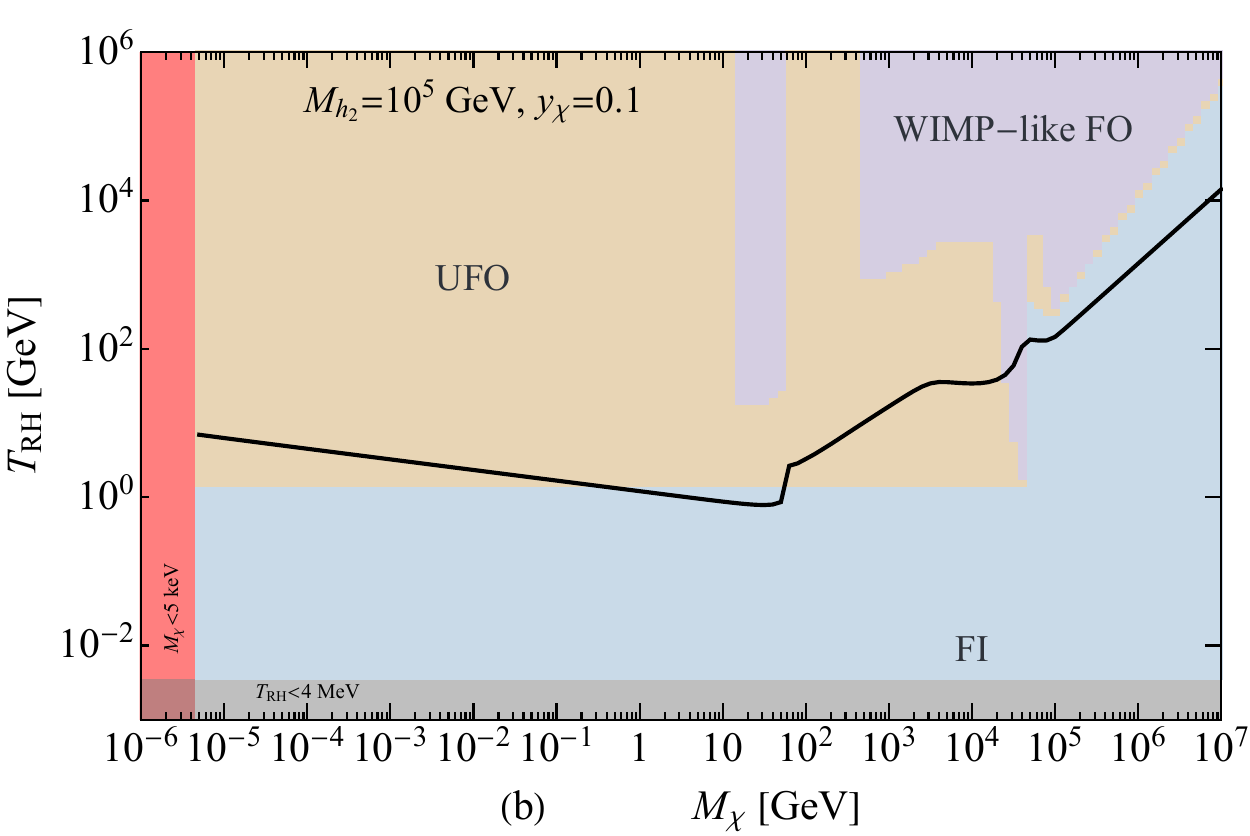}
    \includegraphics[width=0.49\textwidth]{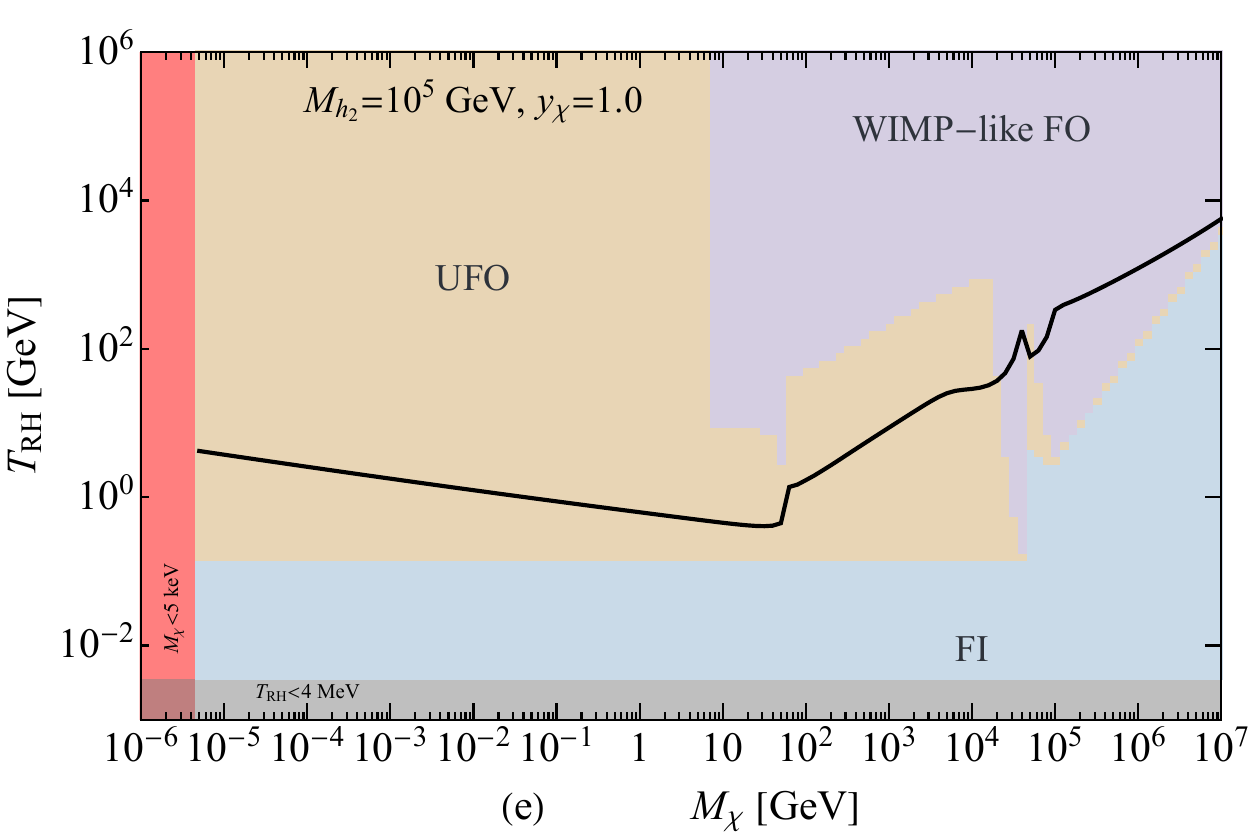}
    \includegraphics[width=0.49\textwidth]{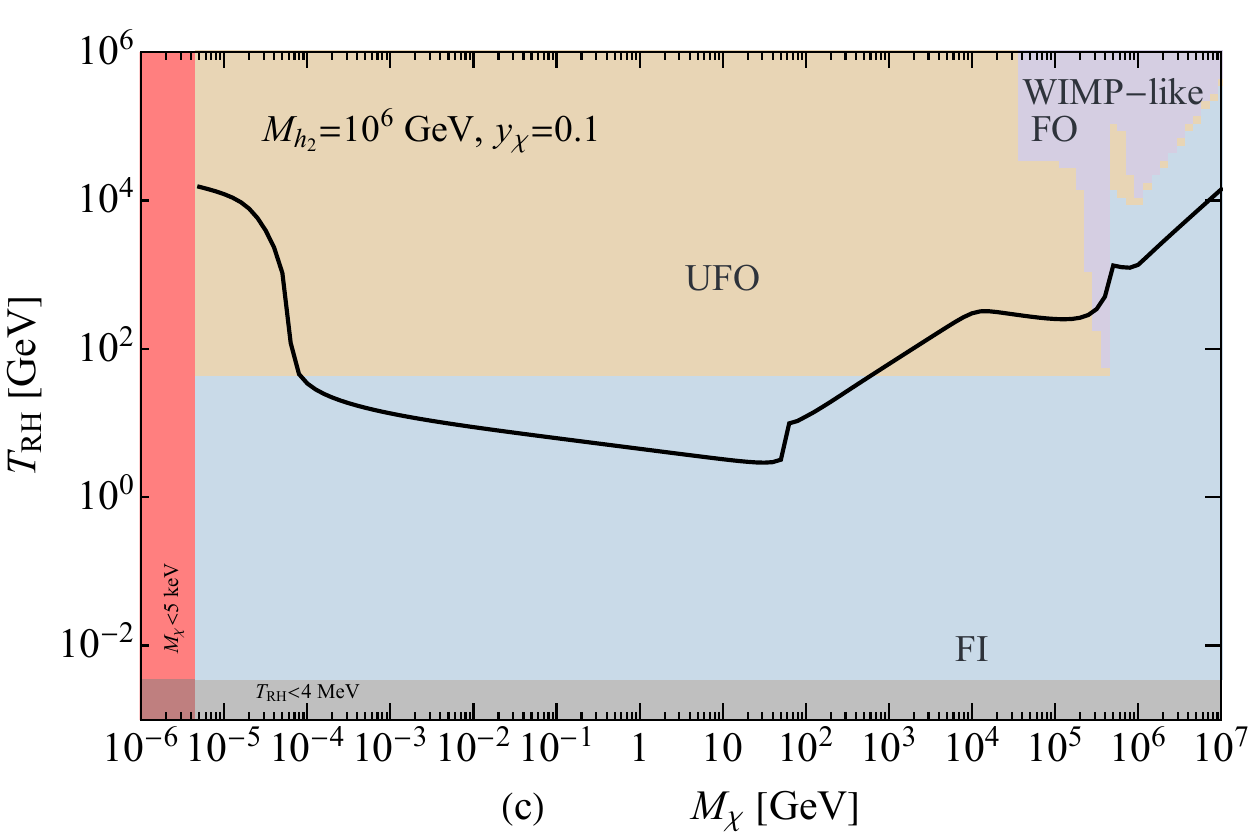}
    \includegraphics[width=0.49\textwidth]{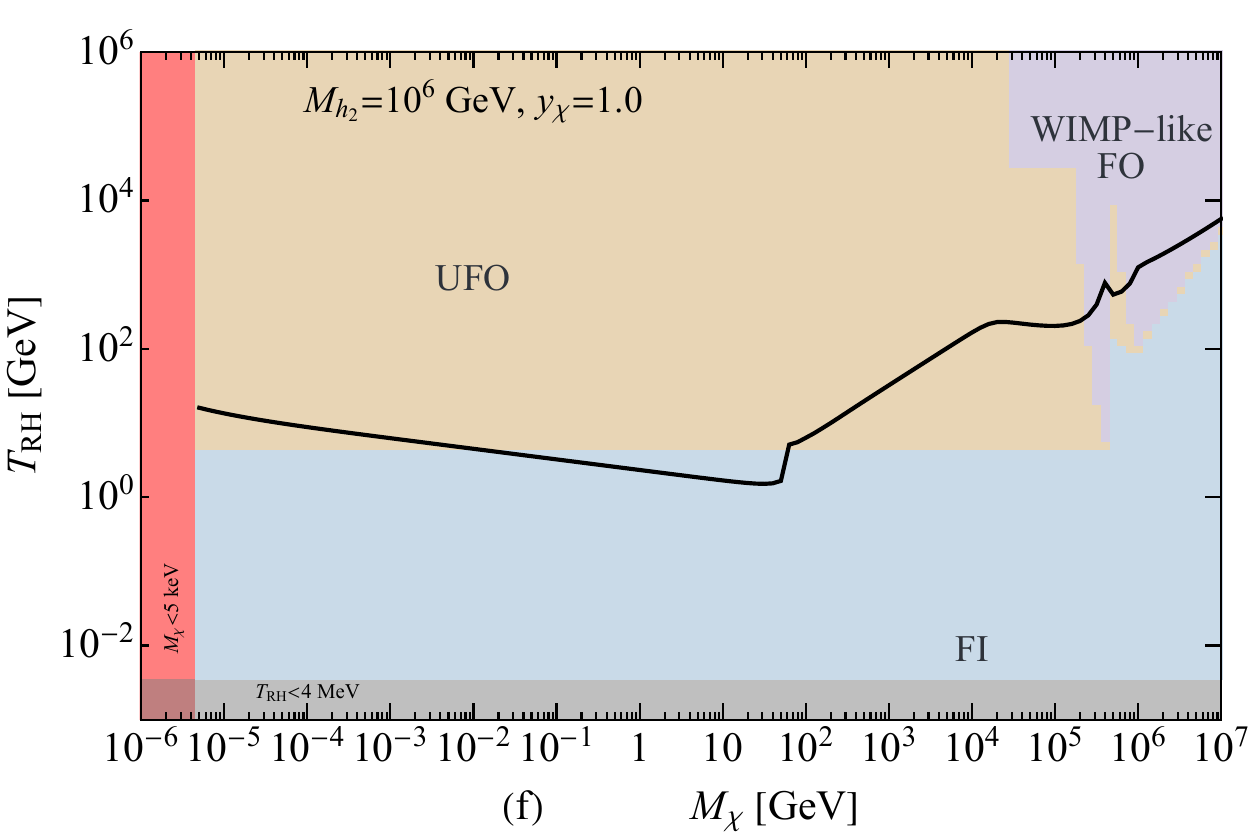}
    \caption{DM production regimes in the $(M_\chi,T_{\mathrm{RH}})$ plane. The mediator masses are $M_{h_2}=10^4$, $10^5$, and $10^6~\mathrm{GeV}$ from top to bottom. Panels (a)--(c) correspond to $y_\chi=0.1$, and panels (d)--(f) to $y_\chi=1$. The light blue, sand, and lavender regions denote FI, UFO, and WIMP-like FO, respectively. The black curves satisfy $\Omega_\chi h^2=0.12$. The gray strip marks $T_{\mathrm{RH}}<4~\mathrm{MeV}$, and the red strip indicates the reference mass cut $M_\chi<5~\mathrm{keV}$.}
\label{fig:relic_parameter_space}
\end{figure*}

Fig.~\ref{fig:relic_parameter_space} shows the relic-density contours and production regimes in the $(M_\chi,T_{\mathrm{RH}})$ plane. The black curves correspond to $\Omega_\chi h^2=0.12$, while the background colors indicate the production mechanism. These production regimes are assigned using the operational equilibrium-tracking and decoupling criteria specified above. Near the regime boundaries, particularly for histories with multiple passages through the vicinity of equilibrium, the classification can depend on the tracking tolerance and the minimum duration requirement. The relic abundance is obtained from the full Boltzmann evolution and is unaffected by the assignment of the regime labels.

A substantial portion of the relic abundance contours lies in the UFO region, especially for a Yukawa coupling as large as $y_\chi=1$. The red shaded region marks $M_\chi<5~\mathrm{keV}$, corresponding to an adopted reference mass cut motivated by structure-formation constraints~\cite{Irsic:2017ixq}. The gray shaded region marks $T_{\mathrm{RH}}<4~\mathrm{MeV}$, which is excluded by the lower limit on the reheating temperature set by big bang nucleosynthesis~\cite{Kawasaki:2000en,Hannestad:2004px,deSalas:2015glj,Hasegawa:2019jsa}. These results demonstrate that the scalar-mediated fermion DM can reproduce the observed relic abundance through distinct thermal histories.

The role of reheating is evident from the dependence of the production regimes on $T_{\mathrm{RH}}$. At a fixed bath temperature $T>T_{\mathrm{RH}}$, the expansion rate during reheating scales as $H\propto T^4/T_{\mathrm{RH}}^2$. Lowering $T_{\mathrm{RH}}$ therefore increases the expansion rate and makes chemical equilibration more difficult, explaining why FI occupies the low-$T_{\mathrm{RH}}$ regions of the panels. For particles produced at a given temperature during reheating, a lower $T_{\mathrm{RH}}$ also leads to stronger dilution by subsequent entropy production. At higher reheating temperatures, interactions can establish chemical equilibrium, and the timing of the final departure from equilibrium determines whether the history is classified as UFO or WIMP-like FO.

Comparing the two columns isolates the effect of the Yukawa coupling. Increasing $y_\chi$ from $0.1$ to $1$ allows chemical equilibrium to be established at lower reheating temperatures, shifting part of the FI region into the UFO regime. For $M_{h_2}=10^6~\mathrm{GeV}$, both Figs.~\ref{fig:relic_parameter_space} (c) and (f) contain FI and UFO segments along the observed relic-density contour. Increasing $y_\chi$ shifts the boundary between these regimes toward lower reheating temperatures and extends the UFO segment on the light-mass branch. Nevertheless, part of the contour remains in the FI region even for $y_\chi=1$. The effect of increasing the coupling therefore depends on both the DM mass and the reheating temperature, rather than producing a uniform change of production mechanism along the entire contour.

On the light-mass branches of the black curves, reproducing the fixed relic density $\Omega_\chi h^2=0.12$ requires a larger final entropy-normalized abundance as $M_\chi$ decreases, since $\Omega_\chi h^2\propto M_\chi\mathcal{Y}_{\chi,\infty}$. For the benchmarks shown, this requirement is generally achieved by increasing $T_{\mathrm{RH}}$. At larger $M_\chi$, production becomes increasingly sensitive to the DM threshold. Scalar mixing introduces a production threshold at $M_\chi=M_{h_1}/2$. As this decay channel becomes phase-space suppressed and eventually closes, the relative importance of the remaining scattering and heavy-scalar decay contributions changes. This accounts for the pronounced change in the relic abundance contours near the light-scalar threshold. And there is similar pronounced change in the relic abundance contours near the heavy-scalar threshold. 

\begin{figure*}[t!]
\centering
    \includegraphics[width=0.49\textwidth]{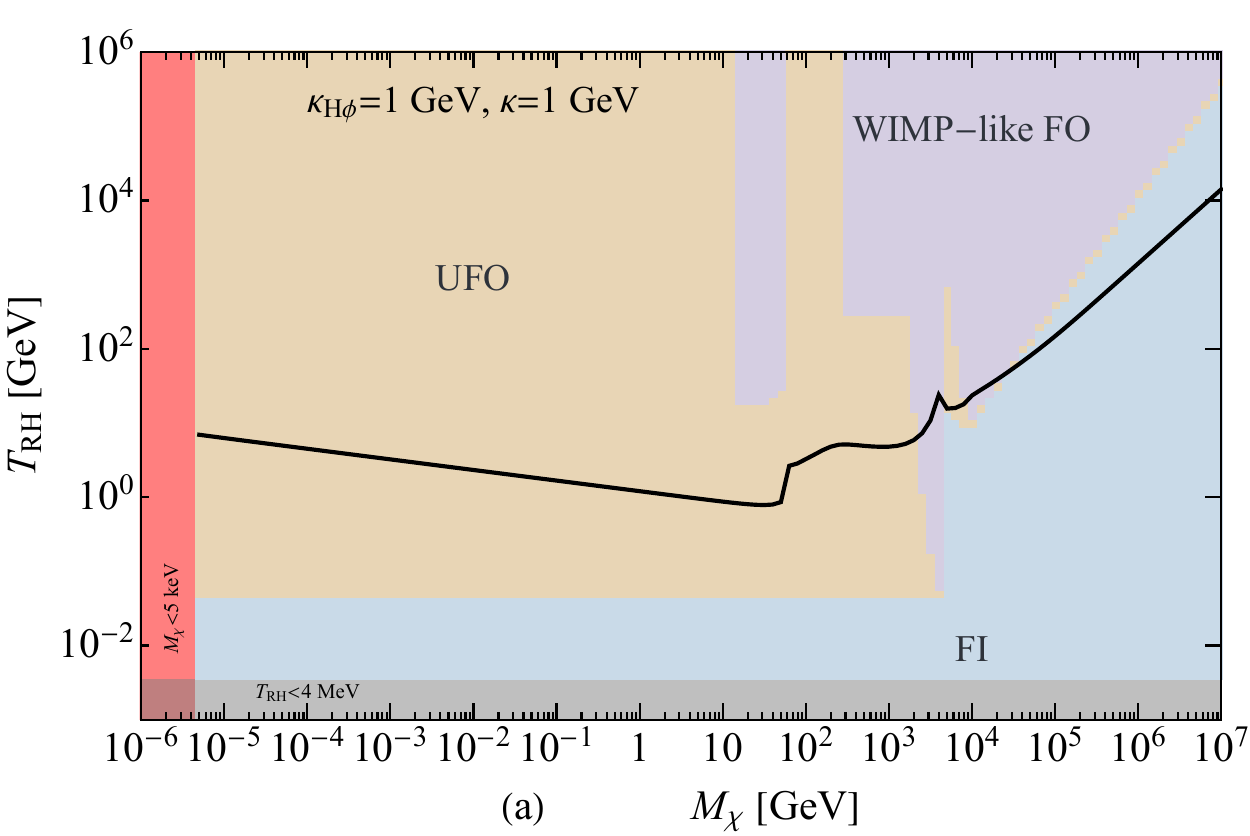}
    \includegraphics[width=0.49\textwidth]{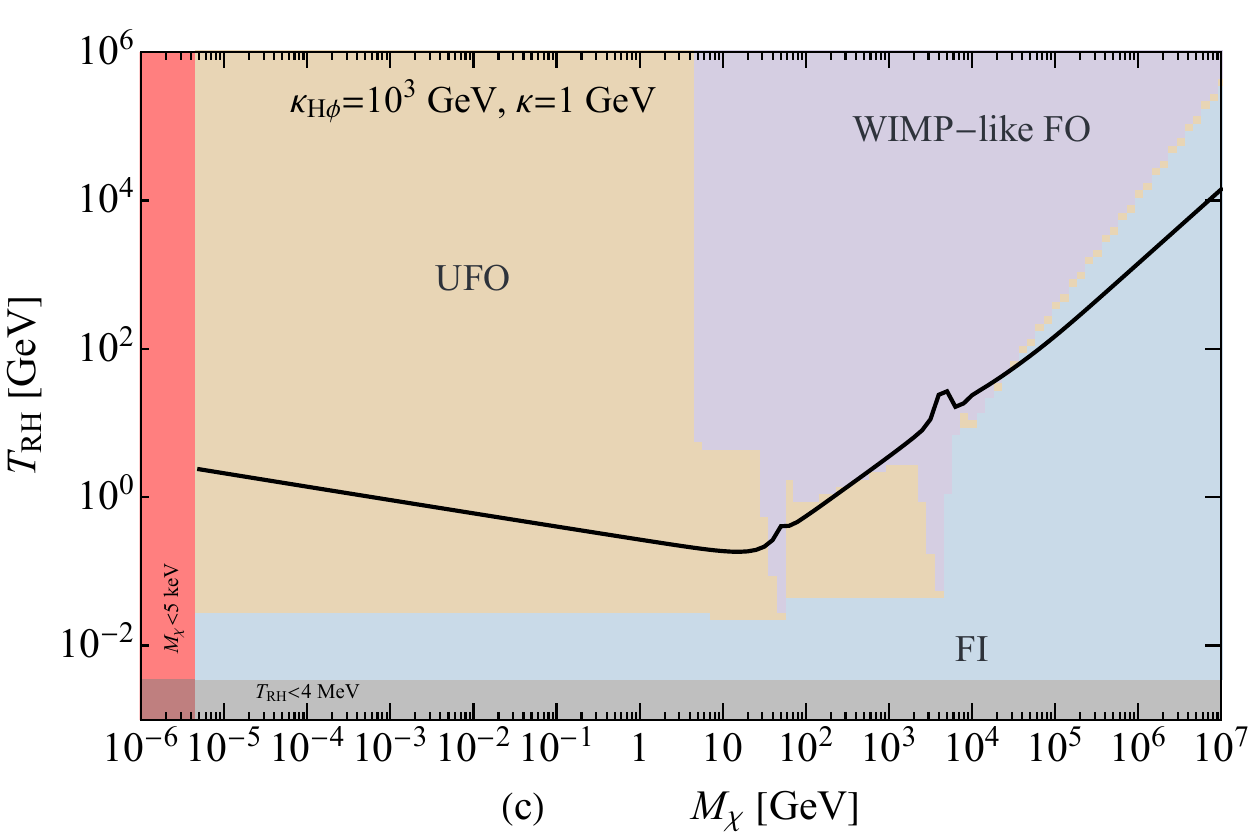}
    \includegraphics[width=0.49\textwidth]{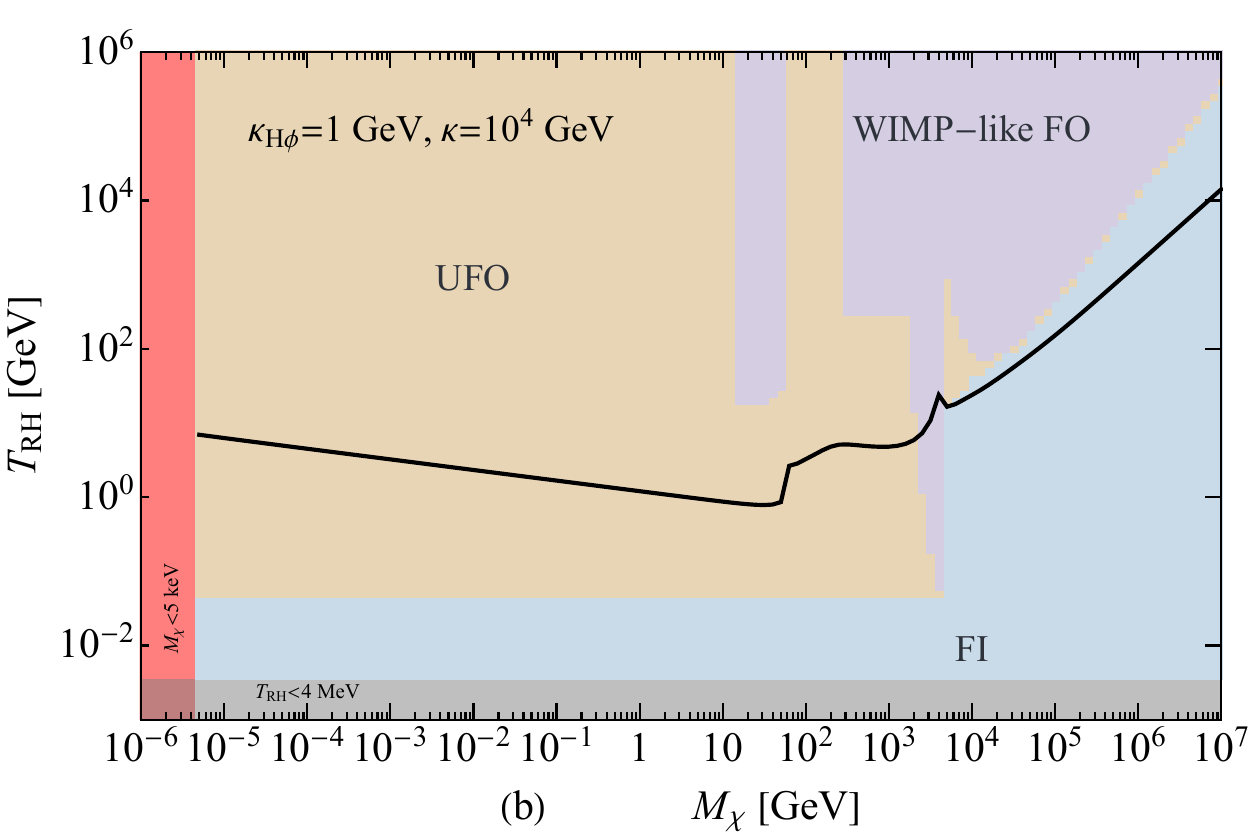}
    \includegraphics[width=0.49\textwidth]{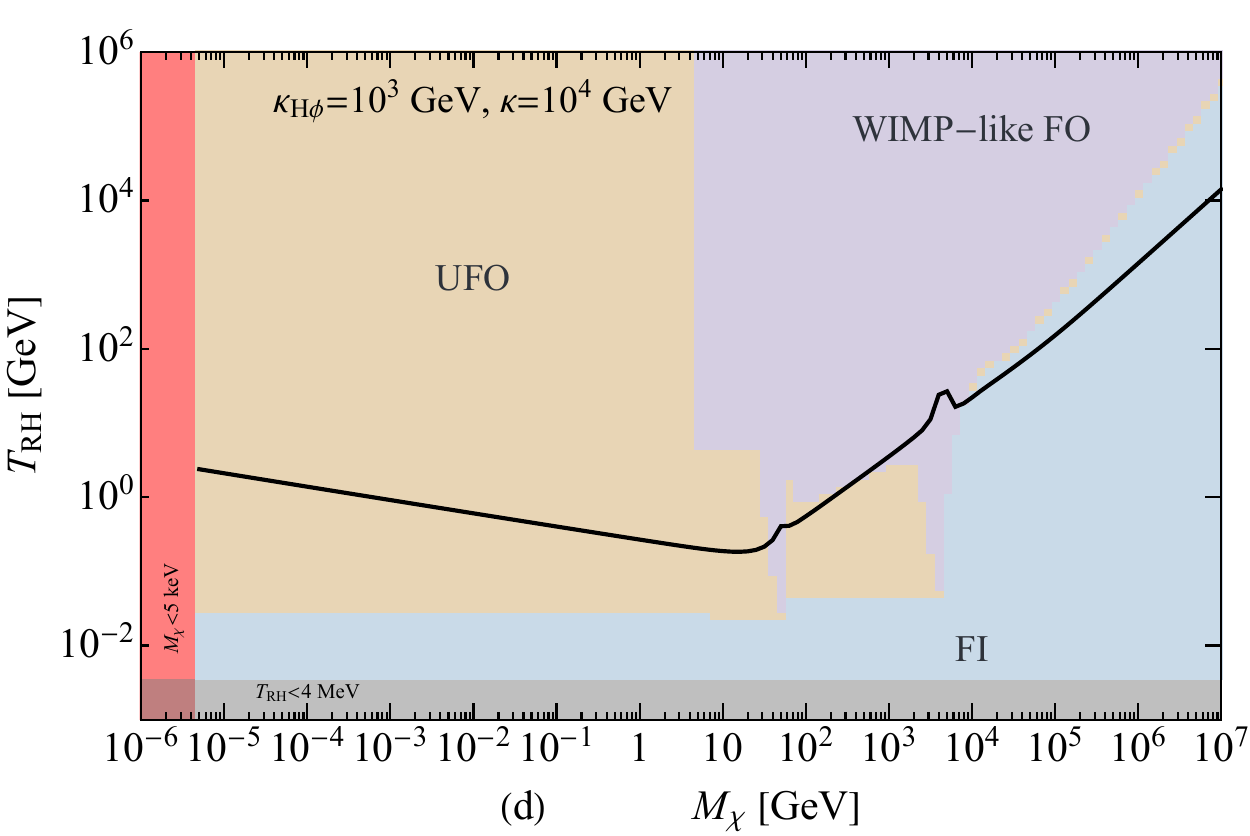}
    \caption{Dependence of the production regimes and relic-density contours on the scalar cubic couplings, for $M_{h_2}=10^4~\mathrm{GeV}$ and $y_\chi=0.1$. The left column has $\kappa_{H\phi}=1~\mathrm{GeV}$, while the right column has $\kappa_{H\phi}=10^3~\mathrm{GeV}$. From top to bottom, $\kappa=1$ and $10^4~\mathrm{GeV}$. The background colors, black relic-density contours, and reference cuts follow the conventions of Fig.~\ref{fig:relic_parameter_space}.}
\label{fig:scalar_coupling_scan}
\end{figure*}

Different from the case of $Z'$ portal DM analyzed in Ref.~\cite{Henrich:2025pca}, the scalar cubic couplings can also affect the production regimes and the relic-density contours. We take $M_{h_2}=10^4\;{\rm GeV},\;y_\chi = 0.1$ and plot Fig.~\ref{fig:scalar_coupling_scan} to examine their effects. The left and right columns correspond to $\kappa_{H\phi} = 1$ GeV and $10^3$ GeV, respectively, while $\kappa = 1$, and $10^4$ GeV from top to bottom. This comparison separates the effects of the interaction connecting the heavy mediator to the light scalar from those of the heavy-scalar self-interaction.

Comparing the two columns shows that increasing $\kappa_{H\phi}$ extends the WIMP-like FO region toward smaller DM masses and lower reheating temperatures. At fixed scalar masses, $\kappa_{H\phi}$ controls both the scalar interactions and the mixing angle. Therefore, it not only affect the contributions from $h_1h_1\to\chi\chi$ illustrated by Eqs.~\eqref{s_channel_amplitude} and \eqref{coupling_constants}, but also affects the DM production through $W^+W^-$, $ZZ$, and $t\bar t$ annihilation, as well as $h_1\to\chi\chi$ when $M_\chi<M_{h_1}/2$. These processes can remain effective after the thermal heavy-scalar population becomes suppressed. Their combined contribution can sustain chemical equilibrium to lower temperatures, shifting part of the UFO region into the WIMP-like FO regime.

The coupling $\kappa_{H\phi}$ also affects the reheating temperature required to reproduce the observed relic abundance. Along the light-mass branches, increasing $\kappa_{H\phi}$ shifts the black contours toward lower $T_{\mathrm{RH}}$. The enhanced interactions with the SM bath affect both equilibrium tracking and out-of-equilibrium production. The observed relic abundance can therefore be obtained at lower $T_{\rm RH}$. In addition, the low-mass FI--UFO boundary moves less than the relic-density contours. This is consistent with thermalization being controlled mainly by heavy-scalar processes at higher temperatures, while production from the SM bath has a stronger influence on the later evolution. The onset of chemical equilibrium and the final relic abundance thus need not have the same sensitivity to $\kappa_{H\phi}$.

Comparing Figs.~\ref{fig:scalar_coupling_scan} (a) and (b), or (c) and (d), we can see that varying $\kappa$ from $1$ to $10^4~\mathrm{GeV}$ produces only small changes on the UFO, WIMP-like FO and FI bounds, as well as the relic-density contours. With scalar mixing retained, $\kappa$ enters the cubic couplings of the scalar mass eigenstates and hence the scalar-exchange amplitudes and their interference with fermion exchange. At fixed scalar masses, $\kappa_{H\phi}$ and $y_\chi$, the Yukawa couplings entering the fermion-exchange amplitudes remain unchanged. At sufficiently high scattering energies, the contribution involving the cubic coupling $\kappa$ is suppressed relative to fermion exchange.


\section{Discussion and Conclusions}
\label{sec:conclusions}
We have studied the production of Majorana fermion DM through a heavy scalar mediator during reheating and the subsequent radiation-dominated era. Our calculation includes mediator decays and scalar scattering processes, retaining the fermion-exchange $t$- and $u$-channel contributions and the relevant interference terms. By solving the full Boltzmann equation and examining the equilibrium-tracking and decoupling history, we identify the parameter regions corresponding to FI, WIMP-like FO, and UFO. All three mechanisms can reproduce the observed DM relic abundance within the parameter space explored. In particular, UFO is realized over extended regions of parameter space, including for Yukawa couplings as large as $y_\chi=1$.

The relative importance of the production channels changes as the universe cools. For the considered benchmarks, $h_2h_2\to\chi\chi$ dominates at sufficiently high temperatures. As the heavy-scalar population becomes Boltzmann suppressed, its contribution to the production rate decreases more rapidly than that from mediator decays. The interactions between DM and SM bath can remain effective at lower temperatures through off-shell mediator exchange. Once DM has thermalized, the suppression of the heavy-particle channels can lead to relativistic decoupling if the remaining DM-SM bath interactions are insufficient to sustain chemical equilibrium.

The parameter scans illustrate how the production channels and reheating history jointly determine the DM abundance and production regime. Lower reheating temperatures make chemical equilibration more difficult and enhance the entropy dilution of DM produced during reheating. The mediator mass governs both the suppression of off-shell scattering and the temperature scales at which the heavy-particle channels become ineffective. At fixed mediator mass and Yukawa coupling, the scalar cubic couplings further modify the relative contributions of the production channels. Increasing $\kappa_{H\phi}$ enhances production through interactions with the SM bath, and can extend equilibrium tracking to lower temperatures. By comparison, the dependence on the heavy-scalar cubic self-coupling $\kappa$ remains modest in the chosen parameter space.


\begin{acknowledgments}
This work has been supported by the National Natural Science Foundation of China (NNSFC) with Grants No. 12075074, No. 12235008, the Hebei Natural Science Foundation for Distinguished Young Scholars with Grant No. A2023201041 and the youth top-notch talent support program of the Hebei Province. 
\end{acknowledgments}

\appendix

\section{Production amplitudes and reaction densities with scalar mixing}
\label{app:fullprod}

\begin{figure}[t]
\centering
\includegraphics[width=\linewidth]{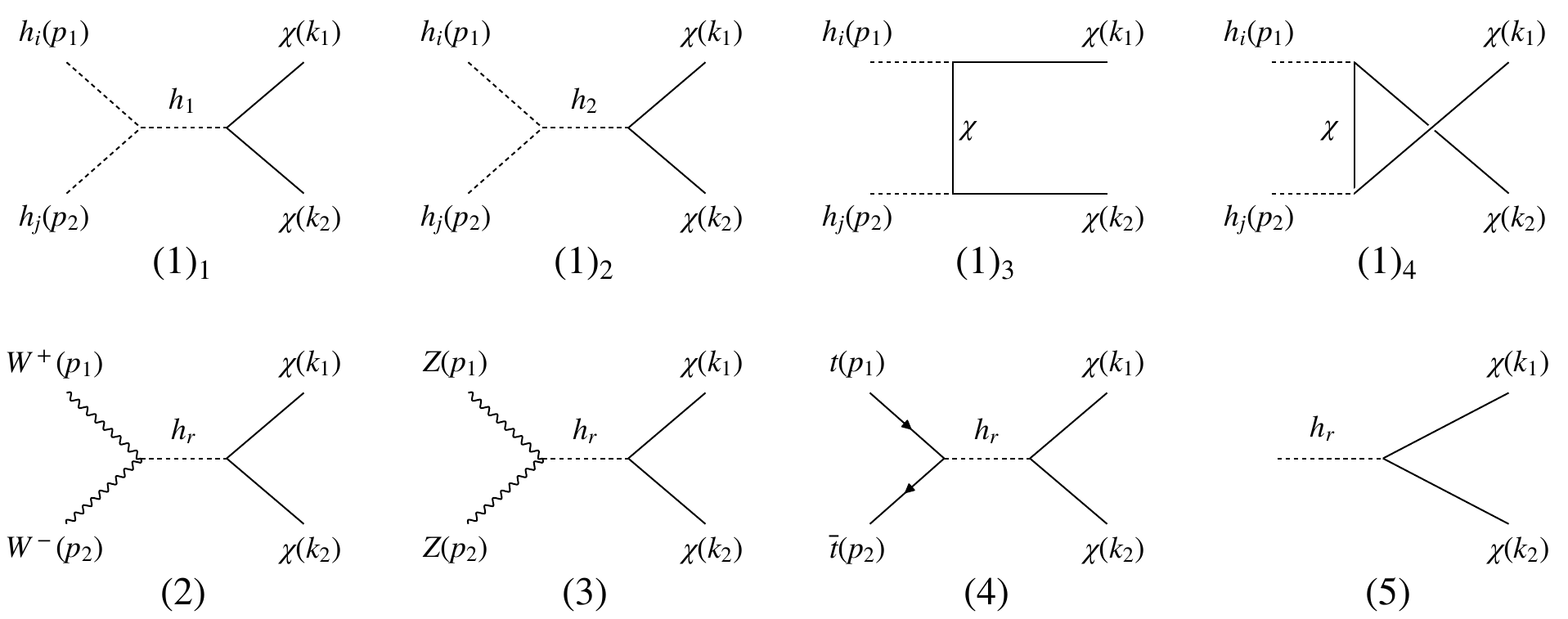}
\caption{Feynman diagrams for the DM production processes included with
nonzero scalar mixing. Panels $(1)_{1,2,3,4}$ represent
$h_i h_j\to\chi\chi$ with $(i,j)=(1,1),(2,2),(1,2)$. Panels (2), (3), and (4) represent $W^+W^-$, $ZZ$, $t\bar t$ annihilation through $h_r\;(r=1,2)$, and (5) represents $h_r\to\chi\chi$ for each kinematically allowed decay.}
\label{app:fullprod:fig}
\end{figure}

This appendix collects the amplitudes and reaction densities used when scalar mixing is retained. The dominant scattering channels are
\begin{equation}
\{h_1h_1,\ h_2h_2,\ h_1h_2,\ W^+W^-,\ ZZ,\ \bar tt \}\to\chi\chi,
\label{app:fullprod:channels}
\end{equation}
together with the decays $h_i\to\chi\chi$, $i=1,2$.  The Feynman diagrams in Fig.~\ref{app:fullprod:fig} constitute the contributions to this set of processes.

The relevant interactions can then be written as
\begin{align}
\mathcal L_{\rm int}\supset{}&\frac{1}{2} y_\chi\sin\theta_{\rm mix}h_1\bar\chi\chi-\frac{1}{2} y_\chi\cos\theta_{\rm mix}h_2\bar\chi\chi
- \frac{1}{3!}\sum_{i,j,k}C_{h_ih_jh_k}h_i h_j h_k\nonumber\\
&+\cos\theta_{\rm mix} h_1\left(\frac{2M_W^2}{v}W^+_\mu W^{-\mu}+\frac{M_Z^2}{v}Z_\mu Z^\mu-\frac{\sqrt2 m_t}{v}\bar t t\right)\nonumber\\
&+\sin\theta_{\rm mix} h_2\left(\frac{2M_W^2}{v}W^+_\mu W^{-\mu}+\frac{M_Z^2}{v}Z_\mu Z^\mu-\frac{m_t}{v}\bar t t\right),
\label{app:fullprod:interactions}
\end{align}
where $C_{h_ih_jh_k}$ is the interaction vertex of $h_ih_jh_k$. For convenience, we also define 
\begin{align}
\mathcal P(s)\equiv\frac{1}{s-M_{h_2}^2+iM_{h_2}\Gamma_{h_2}}-\frac{1}{s-M_{h_1}^2+iM_{h_1}\Gamma_{h_1}},
\label{app:fullprod:propagators}
\end{align}
where $\Gamma_{h_r}\;(r=1,2)$ denotes the total widths of $h_r$.

\subsection{Scattering amplitudes and cross sections}
\label{app:fullprod:scattering}

For a scattering process $a(p_1)b(p_2)\to\chi(k_1)\chi(k_2)$, we use the Mandelstam variables
\begin{align}
s=(p_1+p_2)^2,\quad t=(p_1-k_1)^2,\quad u=(p_1-k_2)^2,
\quad s+t+u=m_a^2+m_b^2+2M_\chi^2.
\label{app:fullprod:stu}
\end{align}
For $h_i h_j\to\chi\chi$, with $(i,j)=(1,1),(2,2),(1,2)$, the two scalar-exchange amplitudes are
\begin{align}
\mathcal M_{h_i h_j}^{(s,h_1)}
&=-y_\chi\sin\theta_{\rm mix}\,
C_{h_i h_j h_1}\frac{1}{s-M_{h_1}^2+iM_{h_1}\Gamma_{h_1}}\,\bar u(k_2)v(k_1),
\label{app:fullprod:scalar_s1}\\
\mathcal M_{h_i h_j}^{(s,h_2)}
&=y_\chi\cos\theta_{\rm mix}\,
C_{h_i h_j h_2}\frac{1}{s-M_{h_2}^2+iM_{h_2}\Gamma_{h_2}}\,\bar u(k_2)v(k_1).
\label{app:fullprod:scalar_s}
\end{align}
The fermion-exchange contributions, expressed with the same ordering of the external Majorana spinors and reduced using their on-shell equations, are
\begin{align}
\mathcal M_{h_i h_j}^{(t)}&=\frac{g_{ij}}{t-M_\chi^2}\bar u(k_2)\bigl(\not\!p_1+2M_\chi\bigr)v(k_1),\label{app:fullprod:scalar_t}\\
\mathcal M_{h_i h_j}^{(u)}&=\frac{g_{ij}}{u-M_\chi^2}\bar u(k_2)\bigl(-\not\!p_1+2M_\chi\bigr)v(k_1),\label{app:fullprod:scalar_u}
\end{align}
where
\begin{align}
g_{ij}=y_\chi^2
\begin{cases}
\sin^2\theta_{\rm mix}, &(i,j)=(1,1),\\
\cos^2\theta_{\rm mix}, &(i,j)=(2,2),\\
-\sin\theta_{\rm mix}\cos\theta_{\rm mix}, &(i,j)=(1,2).
\end{cases}
\label{app:fullprod:yukawa_products}
\end{align}
The full amplitude is therefore
\begin{equation}
\mathcal M_{h_i h_j}
=\mathcal M_{h_i h_j}^{(s,h_1)}
+\mathcal M_{h_i h_j}^{(s,h_2)}
+\mathcal M_{h_i h_j}^{(t)}
+\mathcal M_{h_i h_j}^{(u)}.
\label{app:fullprod:scalar_total}
\end{equation}
The relative signs in these expressions already account for the chosen Majorana spinor ordering. An overall common phase is omitted.

For the dominant SM initial states, the amplitudes following from the specified interactions are
\begin{align}
&\mathcal M_{WW}=\frac{2M_W^2}{v}y_\chi \sin\theta_{\rm mix}\cos\theta_{\rm mix}\mathcal{P}(s)\bigl[\epsilon_+(p_1)\cdot\epsilon_-(p_2)\bigr]\bar u(k_2)v(k_1),
\label{app:fullprod:WW}\\
&\mathcal M_{ZZ}=\frac{2M_Z^2}{v}y_\chi \sin\theta_{\rm mix}\cos\theta_{\rm mix}\mathcal P(s)\bigl[\epsilon(p_1)\cdot\epsilon(p_2)\bigr]\bar u(k_2)v(k_1),
\label{app:fullprod:ZZ}\\
&\mathcal M_{t\bar t}=\frac{m_t}{v}y_\chi \sin\theta_{\rm mix}\cos\theta_{\rm mix}\delta_{\alpha\beta}\mathcal P(s)\bigl[\bar v_t(p_2)u_t(p_1)\bigr]\bigl[\bar u(k_2)v(k_1)\bigr].
\label{app:fullprod:tt}
\end{align}
Here $WW$ denotes $W^+W^-$, and $\alpha,\beta$ are the incoming color indices. For each initial state, the cross section is
\begin{equation}
\sigma_{ab}^{\rm full}(s)=\frac{1}{64\pi s}
\frac{\lambda^{1/2}(s,M_\chi^2,M_\chi^2)}
{\lambda^{1/2}(s,m_a^2,m_b^2)}
\int_{-1}^{1}dz\,
\overline{\left|\mathcal M_{ab}(s,z)\right|^2},
\label{app:fullprod:crosssection}
\end{equation}
where the squared amplitude is defined by
\begin{align}
\overline{|\mathcal M_{ab}|^2}
&\equiv\frac{1}{g_a g_b}
\sum_{\text{all spins, polarizations, and colors}}|\mathcal M_{ab}|^2,
\nonumber\\
g_{h_i}&=1,\qquad g_{W^\pm}=g_Z=3,\qquad g_t=g_{\bar t}=6.
\label{app:fullprod:average}
\end{align}
The factor $1/2!$ for the identical final-state Majorana fermions is already included in Eq.~\eqref{app:fullprod:crosssection}, and the symmetry factor for identical incoming particles is included only in the reaction density below.

\subsection{Decay amplitudes and widths}
\label{app:fullprod:decays}

The amplitudes of $h_1\to\chi\chi$, $h_2\to\chi\chi$ can be written respectively as
\begin{align}
\mathcal M_{h_1}&=-y_\chi\sin\theta_{\rm mix}\,\bar u(k_2)v(k_1),\nonumber\\
\mathcal M_{h_2}&=y_\chi\cos\theta_{\rm mix}\,\bar u(k_2)v(k_1).
\label{app:fullprod:decaychi}
\end{align}
Their partial widths are
\begin{align}
\Gamma_{h_r\to\chi\chi}
&=\frac{1}{2M_{h_r}}\frac{1}{2!}
\int d\Phi_2\,\sum_{\rm spins}|\mathcal M_{h_r}|^2\Theta(M_{h_r}-2M_\chi)\nonumber\\
&=\frac{\lambda^{1/2}(M_{h_r}^2,M_\chi^2,M_\chi^2)}{64\pi M_{h_r}^3}\Theta(M_{h_r}-2M_\chi)\int_{-1}^{1}d\cos\theta\,\sum_{\rm spins}|\mathcal M_{h_r}|^2,
\qquad r=1,2.
\label{app:fullprod:DMwidth}
\end{align}

For the other decay channels entering the total widths and the RIS subtraction, a superscript specifies the final state while the subscript continues to denote the decaying particle:
\begin{align}
&\mathcal M_{h_2\to h_1h_1}=C_{h_1h_1h_2},
\label{app:fullprod:decayhh}\\
&\mathcal M_{h_2\to WW}=\sin\theta_{\rm mix}\frac{2M_W^2}{v}\,\epsilon_+^*(q_1)\cdot\epsilon_-^*(q_2),
\label{app:fullprod:decayWW}\\
&\mathcal M_{h_2\to ZZ}=\sin\theta_{\rm mix}\frac{2M_Z^2}{v}\,\epsilon^*(q_1)\cdot\epsilon^*(q_2),
\label{app:fullprod:decayZZ}\\
&\mathcal M_{h_2\to t\bar t}=\sin\theta_{\rm mix}\frac{m_t}{v}\,\delta_{\alpha\beta}\bar u_t(q_1)v_t(q_2).
\label{app:fullprod:decaytt}
\end{align}
For these channels the partial width is
\begin{align}
\Gamma_{h_2\to ab}
&=\frac{1}{2M_{h_2}S_{ab}}
\int d\Phi_2\,\sum|\mathcal M_{h_2\to ab}|^2
=\frac{\lambda^{1/2}(M_{h_2}^2,m_a^2,m_b^2)}
{32\pi M_{h_2}^3S_{ab}}
\int_{-1}^{1}dz\,\sum|\mathcal M_{h_2\to ab}|^2.
\label{app:fullprod:width}
\end{align}
The sums run over all final-state internal degrees of freedom. The symmetry factors are $S_{h_1h_1}=S_{ZZ}=2$ and $S_{W^+W^-}=S_{t\bar t}=1$.

The total widths entering are constructed using the same couplings as the scattering amplitudes. For the channels retained in this calculation,
\begin{align}
\Gamma_{h_1}
&=\Gamma_{h_1\to\chi\chi}+\Gamma_{h_1\to\mathrm{SM}},
\label{app:fullprod:totalwidth1}\\
\Gamma_{h_2}
&=\Gamma_{h_2\to\chi\chi}+\Gamma_{h_2\to h_1h_1}
+\Gamma_{h_2\to W^+W^-}+\Gamma_{h_2\to ZZ}
+\Gamma_{h_2\to t\bar t}.
\label{app:fullprod:totalwidth2}
\end{align}
Here $\Gamma_{h_1\to\mathrm{SM}}=\cos^2\theta_{\rm mix}\Gamma_{h\to\mathrm{SM}}^{\rm SM}(M_{h_1})$ denotes the total width of the light scalar, where $\Gamma_{h\to\mathrm{SM}}^{\rm SM}(M_{h_1})$ is the SM Higgs total width. 

\subsection{Reaction densities and the total production rate}
\label{app:fullprod:rates}

In the Maxwell–Boltzmann approximation, the scattering reaction density reads 
\begin{align}
\gamma_{ab}^{\rm full}(T)
&=\frac{g_a g_b\,T}{32\pi^4(1+\delta_{ab})}
\int_{s_{ab}^{\rm min}}^\infty ds\,
\frac{\lambda(s,m_a^2,m_b^2)}{\sqrt{s}}
\sigma_{ab}^{\rm full}(s)
K_1\!\left(\frac{\sqrt{s}}{T}\right),
\nonumber\\
s_{ab}^{\rm min}
&=\max\bigl\{(m_a+m_b)^2,4M_\chi^2\bigr\}.
\label{app:fullprod:reaction_scattering}
\end{align}
The factor $g_a g_b$ restores the initial-state sum averaged in the cross section. Here, $\delta_{ab}=1$ for identical species $h_1h_1$, $h_2h_2$, $ZZ$, and $\delta_{ab}=0$ for $h_1h_2$, $W^+W^-$, and $t\bar t$. The scalar contribution is
\begin{equation}
\gamma_{ij}^{\rm full}(T)
=\frac{T}{32\pi^4(1+\delta_{ij})}
\int_{s_{ij}^{\rm min}}^\infty ds\,
\frac{\lambda(s,M_{h_i}^2,M_{h_j}^2)}{\sqrt{s}}
\sigma_{ij}^{\rm full}(s)
K_1\!\left(\frac{\sqrt{s}}{T}\right).
\label{app:fullprod:reaction_scalar}
\end{equation}

For an equilibrium scalar $h_r$, which has one internal degree of freedom, the decay reaction density is
\begin{align}
\gamma_{D,r}(T)
&=\int\frac{d^3p}{(2\pi)^3}\,e^{-E_r/T}
\frac{M_{h_r}}{E_r}\,\Gamma_{h_r\to\chi\chi}
\nonumber\\
&=\frac{M_{h_r}^2T}{2\pi^2}
K_1\!\left(\frac{M_{h_r}}{T}\right)
\Gamma_{h_r\to\chi\chi},
\qquad E_r=\sqrt{\boldsymbol p^{\,2}+M_{h_r}^2}.
\label{app:fullprod:reaction_decay}
\end{align}
The quantities $\gamma_{ab}$ and $\gamma_{D,r}$ count reactions per unit volume per unit time.

Explicit scalar decays overlap with the on-shell part of resonant scattering. For a narrow and isolated scalar resonance, the real intermediate state contribution is obtained from
\begin{equation}
|P_r(s)|^2\ \longrightarrow\
\frac{\pi}{M_{h_r}\Gamma_{h_r}}\,
\delta(s-M_{h_r}^2).
\label{app:fullprod:NWA}
\end{equation}
Consequently, for an included initial state $ab$,
\begin{align}
\gamma_{ab}^{\mathrm{RIS},r}(T)
&=\frac{M_{h_r}^2T}{2\pi^2}
K_1\!\left(\frac{M_{h_r}}{T}\right)
\frac{\Gamma_{h_r\to ab}\Gamma_{h_r\to\chi\chi}}
{\Gamma_{h_r}}
\nonumber\\
&=\operatorname{Br}(h_r\to ab)\,\gamma_{D,r}(T).
\label{app:fullprod:RIS}
\end{align}
This subtraction is applied only when both partial decays are kinematically allowed and $h_r\to\chi\chi$ is included explicitly. It removes the narrow-width on-shell contribution of the squared resonant propagator, while the full scattering amplitudes retain the interference between diagrams.

Summing over the included initial states
$\mathcal B=\{h_1h_1,h_2h_2,h_1h_2,W^+W^-,ZZ,t\bar t\}$,
the DM particle-number production rate is
\begin{equation}
R_\chi(T)=2\left\{
\sum_{ab\in\mathcal B}\left[
\gamma_{ab}^{\rm full}(T)-\sum_{r=1}^2
\gamma_{ab}^{\mathrm{RIS},r}(T)\right]
+\gamma_{D,1}(T)+\gamma_{D,2}(T)\right\}.
\label{app:fullprod:Rgeneral}
\end{equation}
With$M_{h_1}=125~\mathrm{GeV}$, none of the listed initial states can access the $h_1$ pole. The $h_2h_2$ and $h_1h_2$ initial states also cannot access the $h_2$ pole. Then the total rate becomes
\begin{align}
R_\chi(T)=2\Bigl[
&\gamma_{11}^{\rm full}(T)+\gamma_{22}^{\rm full}(T)
+\gamma_{12}^{\rm full}(T)
\nonumber\\
&+\gamma_{WW}^{\rm full}(T)+\gamma_{ZZ}^{\rm full}(T)
+\gamma_{t\bar t}^{\rm full}(T)
+\gamma_{D,1}(T)+\gamma_{D,2}(T)
\nonumber\\
&-\gamma_{11}^{\mathrm{RIS},2}(T)
-\gamma_{WW}^{\mathrm{RIS},2}(T)
-\gamma_{ZZ}^{\mathrm{RIS},2}(T)
-\gamma_{t\bar t}^{\mathrm{RIS},2}(T)
\Bigr].
\label{app:fullprod:Rexplicit}
\end{align}
Kinematically closed partial decays and their corresponding RIS terms contribute zero. 

\bibliography{Refs}

\end{document}